\def\draftversion{false}

\RequirePackage{ifthen}
\ifthenelse{\equal{\draftversion}{true}}{
	\documentclass[aps,prl,10pt,galley,amsmath,amssymb,
	floatfix,superscriptaddress,
	citeautoscript]{revtex4}
}{
	\documentclass[aps,prl,10pt,twocolumn,amsmath,amssymb,
	longbibliography,
	superscriptaddress,citeautoscript]{revtex4-1}
}

\usepackage{graphicx}
\usepackage[usenames,dvipsnames]{color} 
\usepackage{bm} 
\usepackage[update,prepend]{epstopdf}
\usepackage{soul} 
\usepackage{dcolumn} 
\usepackage{comment}
\usepackage{gensymb}
\usepackage[linktocpage=true]{hyperref}
\usepackage{hyperref}
\usepackage{comment}
\usepackage{notes2bib}
\hypersetup{
  pdfnewwindow=true, colorlinks=true,
  linkcolor=blue, anchorcolor=blue,
  citecolor=blue, filecolor=blue,
  menucolor=blue, urlcolor=blue}

\def\Red#1{\textcolor{red}{#1}}

\def\comment#1{}

\DeclareMathAlphabet\mathbfcal{OMS}{cmsy}{b}{n}

\ifthenelse{\equal{\draftversion}{true}}{
	\marginparwidth 2.7in
	\marginparsep 0.5in
	\newcounter{comm} 
	\def\commnext{\stepcounter{comm}}
	\def\commtext{{\bf\color{OliveGreen}[\arabic{comm}]}}
	\def\commmar{{\bf\color{OliveGreen}[\arabic{comm}]}}
	\def\dvm#1{\commnext\marginpar{\small DV\commmar: #1}\commtext}

	\def\krm#1{\commnext\marginpar{\small KR\commmar: #1}\commtext}

}{
	\def\krm#1{}
    \def\dvm#1{}
	
}

\def\Red#1{\textcolor{red}{#1}}

\begin{document}


\title{Cyclic Ferroelectric Switching and Quantized Charge Transport in CuInP$_2$S$_6$}

\author{Daniel Seleznev}
\affiliation{Department of Physics and Astronomy, Center for Materials Theory, Rutgers University, Piscataway, New Jersey 08854, USA}

\author{Sobhit Singh}
\affiliation{Department of Mechanical Engineering, University of Rochester, Rochester, New York 14627, USA}

\author{John Bonini}
\affiliation{Center for Computational Quantum Physics, Flatiron Institute, 162 5th Avenue, New York, New York 10010, USA}

\author{Karin M. Rabe}
\affiliation{Department of Physics and Astronomy, Center for Materials Theory, Rutgers University, Piscataway, New Jersey 08854, USA}

\author{David Vanderbilt}
\affiliation{Department of Physics and Astronomy, Center for Materials Theory, Rutgers University, Piscataway, New Jersey 08854, USA}

\begin{abstract}

The van der Waals layered ferroelectric CuInP$_2$S$_6$ has been found to exhibit a variety of intriguing properties arising  from the fact that the Cu ions are unusually mobile in this system. While the polarization switching mechanism is usually understood to arise from Cu ion motion within the monolayers, a second switching path involving Cu motion across the van der Waals gaps has been suggested. In this work, we perform zero-temperature first-principles calculations on such switching paths, focusing on two types that preserve the periodicity of the primitive unit cell: ``cooperative" paths preserving the system's glide mirror symmetry, and ``sequential" paths in which the two Cu ions in the unit cell move independently of each other. We find that CuInP$_2$S$_6$ features a rich and varied energy landscape, and that sequential paths are clearly favored energetically both for cross-gap and through-layer paths.  Importantly, these segments can be assembled to comprise a globally insulating cycle with the out-of-plane polarization evolving by a quantum as the Cu ions shift to neighboring layers.  In this sense, we argue that CuInP$_2$S$_6$ embodies the physics of a quantized adiabatic charge pump.

\end{abstract}

\maketitle
CuInP$_2$S$_6$ (CIPS) is a van der Waals (vdW) layered ferroelectric (FE) that has drawn much attention in recent years due to its unique properties and promise for application \cite{xue-apr2020,zhang-natrevmat2022,zhou-fop2020}. Much of the interest in CIPS has been driven by its ability to maintain stable ferroelectricity in the 2D limit \cite{liu-natcomm2016} without resorting to extrinsic mechanisms such as strain \cite{choi-science2004,garcia-nature2009,zhang-prb2014} or compensation charges \cite{sai-prb2005}. However, other claimed or observed intriguing properties, such as negative longitudinal piezoelectricity \cite{you-sciadv2019,brehm-natmat2020,neumayer-prm2019} and negative capacitance \cite{neumayer-aem2020,ohara-aem2022}, have further motivated the investigation of this system.

The Cu ions play a central role in the physics of CIPS. First and foremost, the low-temperature FE phase results from a Cu off-centering instability at $T_c\sim310$ K \cite{maisonneuve-prb1997,vysochanskii-1998}. Furthermore, under applied strain, the ferroelectricity in CIPS has been found to exhibit a quadruple-well potential for out-of-plane (OOP) Cu displacements \cite{brehm-natmat2020}. That is, there are two additional high-polarization (HP) states, associated with the strain-induced appearance of stable Cu positions just inside the vdW gaps. Both the low polarization (LP) and HP states feature large, \r{A}-scale spontaneous polar displacements, and the $+$LP and $+$HP (and $-$LP and $-$HP) states are separated by a small, strain-tunable energy barrier. Finally, at higher temperatures, Cu ionic conductivity is observed, with the onset temperature in fact even below the FE transition \cite{zhou-materhoriz2020,balke-ami2018,maisonneuve-FE1997,banys-pt2004,dziaugys-pt2013,macutkevic-pssb2015}. 

The principal ``up-down" polarization switching mechanism in CIPS involves the movement of Cu ions within the monolayers, corresponding to the off-centering instability mentioned above \cite{balke-ami2018,neumayer-prm2019,brehm-natmat2020,neal-prb2022}. However, the high-temperature ionic conductivity also suggests the possibility of a novel switching pathway involving Cu ion migration across the vdW gaps, as supported by some experimental studies \cite{neumayer-pra2020,neumayer-aem2020,zhou-materhoriz2020,zhang-nl2021}. Previous first-principles theoretical studies first considered a ``synchronous'' cross-gap switching pathway in which the primitive-cell periodicity and equivalence of the layers was preserved \cite{neumayer-aem2020}, although the structural evolution along the switching path was not reported in detail.  Subsequent work by the same group considered ``asynchronous'' mechanisms instead, using an $8\times8$ in-plane supercell geometry to study processes in which Cu ions jump one at a time from a given layer to its neighbor \cite{ohara-aem2022}. The evolution of the OOP polarization was discussed in these works \cite{neumayer-aem2020,ohara-aem2022}, although the interpretation in the synchronous case was later revised \cite{neumayer-aem2021}.

The energy landscape of CIPS is quite complex, allowing for the possibility of a variety of competing paths. In this paper, we take a systematic approach focusing on the identification of synchronous switching pathways. That is, we constrain the system to retain the periodicity of the primitive 20-atom unit cell, and use zero-temperature first-principles density functional theory (DFT) calculations to search for low-energy pathways allowing both Cu atoms to cross the vdW gaps into adjacent layers. The characterization of such pathways is an important preliminary to understanding the practical switching that can occur at higher temperature, which typically occurs via domain wall propagation.

In particular, we find a path in which the Cu ions in both layers move cooperatively involving a rather high energy barrier. Instead, a much lower barrier is achieved by a path that allows for the sequential movement of the Cu ions, a possibility not considered in the earlier work \cite{brehm-natmat2020,neumayer-aem2020,ohara-aem2022}. This path includes multiple saddle points and local minima corresponding to different local Cu chemical environments. We also identify sequential movement as being energetically favorable for through-layer switching. The system remains robustly insulating everywhere along both sequential switching paths, which together form a cycle transporting each Cu ion from one layer to the next. The computed change of Berry-phase polarization along this cycle corresponds to the transport of a single unit of charge by one lattice vector. The overall process is thus a FE switching cycle that embodies a quantized adiabatic charge pump \cite{thouless-prb1983} accomplished by ionic transport \cite{jiang-prl2012}.

\begin{figure}
\centering
\includegraphics[width= 7.6cm]{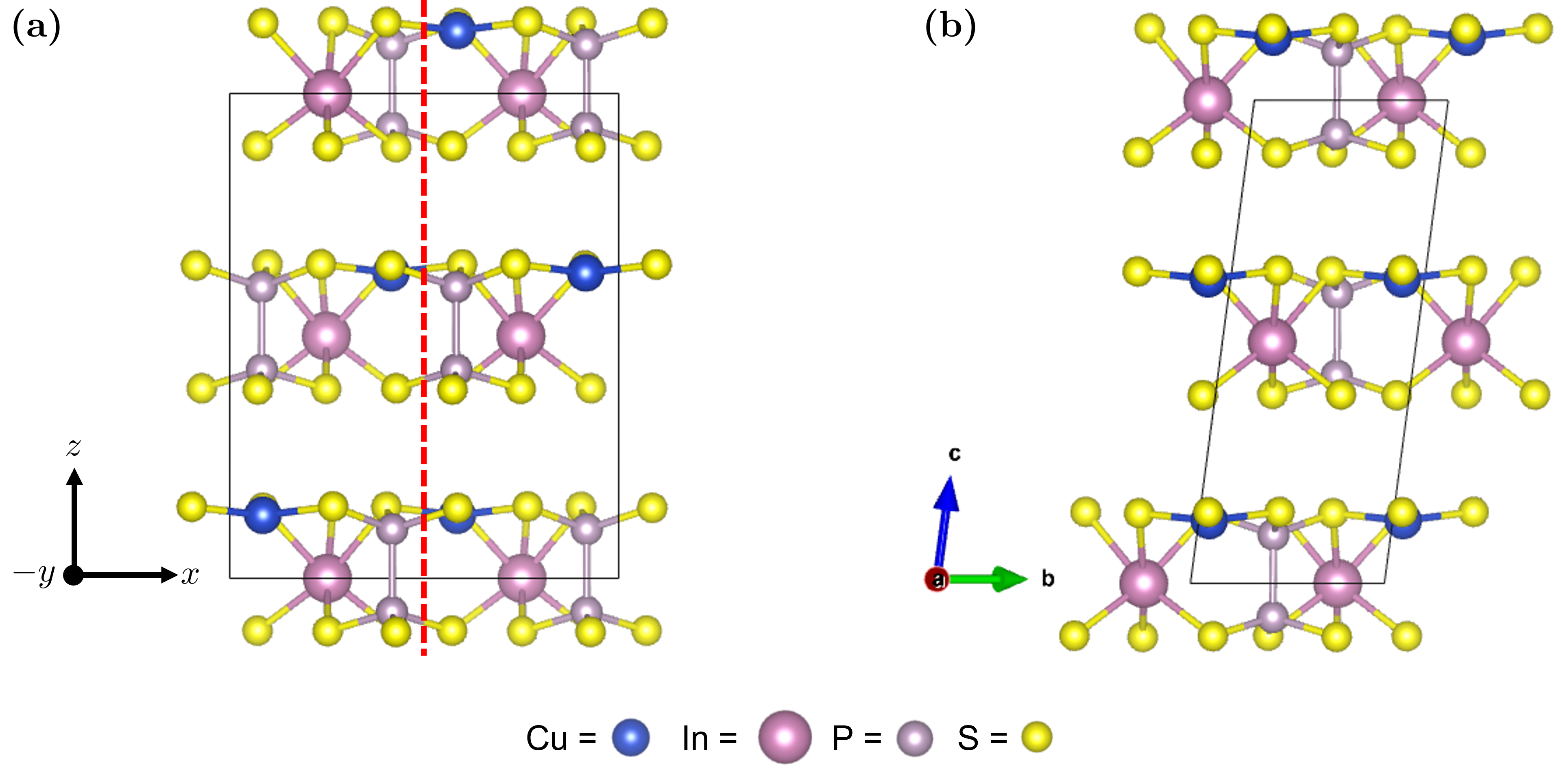}
\caption{(a) $P^+$ polar phase of CIPS in the conventional cell setting. (b) Same, but in the primitive unit cell setting. The Cu ions reside inside the monolayers and above the monolayer midplanes. The dashed red line in (a) indicates the glide mirror plane of the system. The lattice parameters and cell angles are $a=b=6.11$ \r{A}, $c=13.36$ \r{A} and $\alpha=\beta=85.67^{\circ}$, $\gamma=120.0^{\circ}$.}
\label{fig:fig1}
\end{figure}

CIPS is a layered vdW system that is a member of the transition metal thiophosphate family, in which metal cations are found embedded in a lattice framework of (P$_2$S$_6$)$^{4-}$ anions. With two monolayers per primitive unit cell, the individual layers of CIPS are coupled to each other by weak vdW forces, and feature Cu$^{1+}$ and In$^{3+}$ cations surrounded by sulfur octahedra, with P-P dimers filling the octahedral voids (see Fig.~\ref{fig:fig1}). At the Curie temperature $T_c\sim 310$ K, CIPS undergoes a first order order-disorder paraelectric (PE) to FE transition, accompanied by a space-group symmetry reduction from $C2/c\to Cc$ \cite{simon-cm1994,maisonneuve-prb1997}.

Below $T_c$, the nominal centrosymmetric structure is unstable to a Cu off-centering instability \cite{maisonneuve-prb1997,vysochanskii-1998}, and the Cu ions consequently occupy locations above or below the center planes of the monolayers, corresponding to two configurations that we denote as $P^+$ and $P^-$ for ``up'' and ``down'' OOP polarization, respectively \cite{simon-cm1994,maisonneuve-prb1997,maisonneuve-jac1995}. The former is shown in Fig.~\ref{fig:fig1}. However, because we shall consider cross-gap and cyclic evolution shortly, we emphasize that these labels denote the location of a Cu ion relative to its current host layer, and not to the sign of polarization in any global sense.

Importantly, both the high- and low-symmetry monoclinic space groups share a glide mirror plane $\{M_x|\textbf{c}/2\}$, which maps the Cu ion in one layer onto the one in the neighboring layer. Thus, any polarization switching path -- be it involving Cu motion through the monolayers or across the vdW gaps -- that preserves the glide mirror symmetry will involve cooperative motion of both Cu ions.

In the language of the Berry-phase theory \cite{Resta2007}, on such a path the system must pass through a state in which the formal OOP polarization is either zero or half of the quantum. This may occur because the state in question
has some additional symmetry, such as inversion, that enforces these
values. The aforementioned $C2/c$ centrosymmetric structure corresponding to the nominal PE phase is one such example. Other structures within the same space group also occur when both Cu ions reside at inversion centers located in the midplanes of the vdW gaps. For future reference, we also note that $C2/c$ features two subgroups, namely $P\overline{1}$ and $C2$, that while breaking glide mirror symmetry, still ensure a vanishing OOP polarization.

Our search for such cooperative switching paths proceeds as follows. By $\Delta_c$ we denote the difference between the Cu and In internal coordinates along the $\textbf{c}$ lattice vector (note that $\Delta_c$ is defined modulo $1/2$). Starting with the $P^+$ state, where $\Delta_c = 0.13$, the Cu sublattice is first incrementally shifted downward along $\textbf{c}$ toward the nominal centrosymmetric PE structure at $\Delta_c=0$. At each step the lattice vectors and internal coordinates are relaxed within the $Cc$ space group subject to the $\Delta_c$ constraint. This procedure terminates by converging on the expected $C2/c$ structure when $\Delta_c$ reaches zero. The structural energy of the system increases monotonically along this path, reaching $0.72$\,eV for the final $C2/c$ structure, where we have adopted the convention of reporting all structural energies relative to that of the $P^{\pm}$ ground states on a per-unit-cell basis.

The same procedure is then repeated in the opposite direction as $\Delta_c$ is increased toward $0.25$, the nominal value for Cu atoms centered in the vdW gap. The energy again increases monotonically, but this time, we are surprised 
to find that the system remains in the $Cc$ space group up to and including the terminal structure. We label this structure as $M^+$ and find its energy to be $0.97$\,eV. We also perform the same set of calculations starting from $P^-$ at $\Delta_c=-0.13$, stopping this time after $\Delta_c$ is increased to 0 or decreased to $-0.25$. The resulting structures are found to be inversion images of those found above, with opposite sign of $\Delta_c$ and identical energy, with the particular structure at $\Delta_c=-0.25$ denoted as $M^-$. The full energy profile as a function of $\Delta_c$ is depicted in the Supplemental Material (SM) \cite{SM}.  

\begin{figure}
\centering
\includegraphics[width=8.5cm]{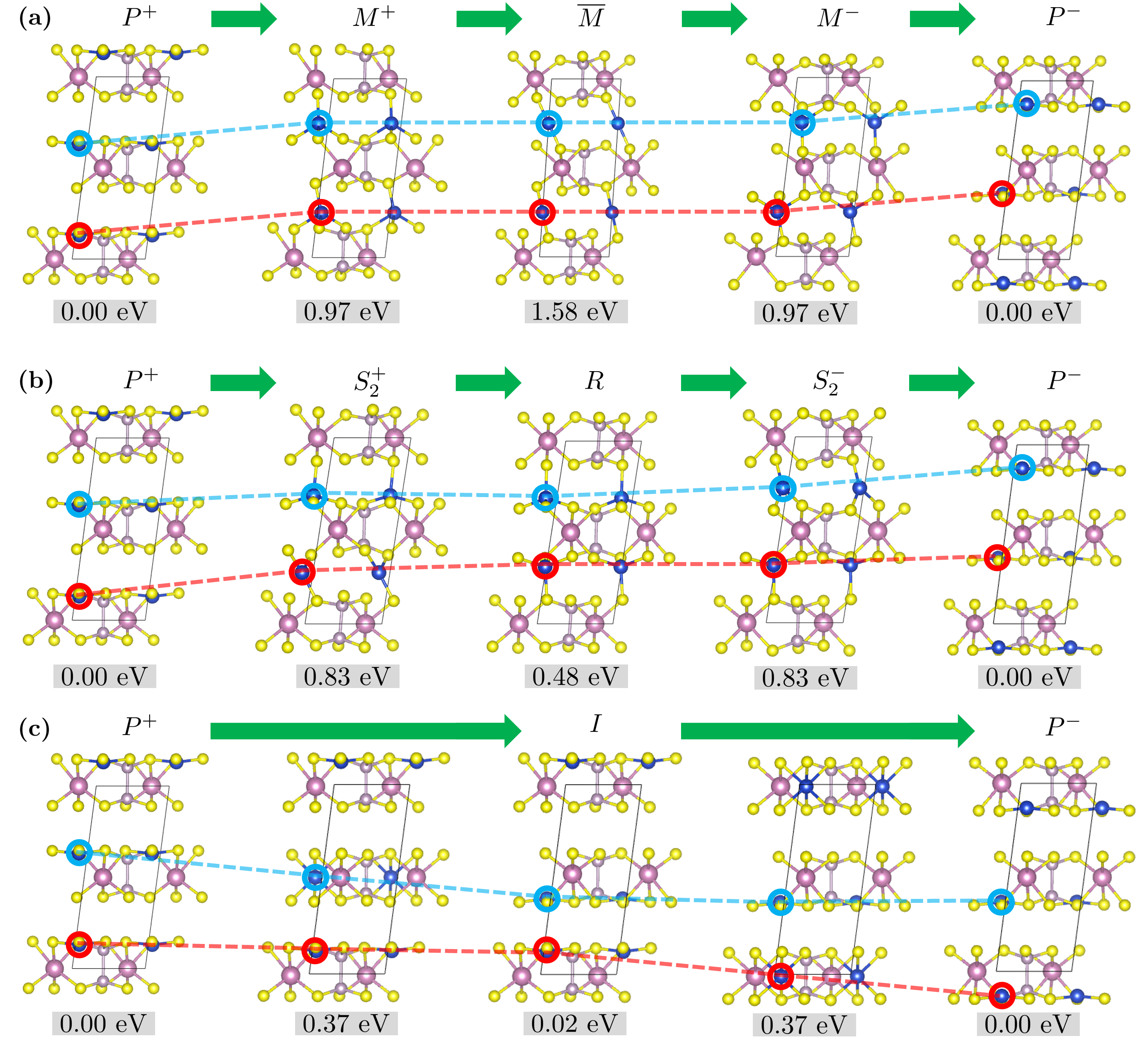}
\caption{(a) The representative sequence $P^+\to M^+\to\overline{M}\to M^-\to P^-$. The Cu ions in the lower (red) and upper (blue) layers of the unit cell move cooperatively along the $\textbf{c}$ lattice vector throughout the switching path, as emphasized by the guiding lines. (b) The representative sequence $P^+\to S^+_2\to R\to S^-_2\to P^-$. The lower Cu ions move across the vdW gap first, and are subsequently followed by the Cu ions in the upper vdW gap. (c) Sequence of structures from $P^+$ to $P^-$ passing through $I$ in the middle. The upper Cu ions move through a monolayer first, followed by the bottom Cu ions. The number below a structure indicates its energy relative to $P^{\pm}$.}
\label{fig:fig2}
\end{figure}

Although $M^-$ and $M^+$ feature identical energies and are inversion images of each other, we crucially find that the structures are distinct, as is evident from the second and fourth panels of Fig.~\ref{fig:fig2}(a).  While the Cu ions see a roughly tetrahedral environment in both cases, three of the four bonds are to S atoms in the lower layer for $M^+$, and the reverse is true for the $M^-$. As a result, a continuous path across the vdW gap is not yet identified.

To complete the cross-gap cooperative switching path, we perform a series of nudged elastic band (NEB) calculations, detailed in the SM \cite{SM}, from which we find that the midpoint of the NEB path, which we designate as $\overline{M}$, is a structure with $C2/c$ symmetry. The full sequence $P^+\to M^+\to\overline{M}\to M^-\to P^-$ is illustrated in the five panels of Fig.~\ref{fig:fig2}(a), highlighting the cooperative nature of Cu motion along the $\textbf{c}$ lattice vector. In passing along this path, the Cu ions visit a variety of coordination environments. The Cu ions begin in a nearly trigonal planar setting ($P^+)$, then pass through the distorted tetrahedral coordination environment ($M^+)$ mentioned above, before reaching the midpoint of the switching path ($\overline{M}$) where a linear two-fold coordination appears. On the second half of the path, the Cu ions pass through the same coordination environments but in reverse. The energy barrier for the switching path is defined by $\overline{M}$, and is found to be 1.58\,eV.  Such a high barrier suggests that this switching path is not very plausible, and that alternative paths should be investigated.

The unstable zone center phonon modes of the $\overline{M}$ structure provide some insights that motivate another strategy for investigating lower energy switching paths. Not surprisingly, we find one unstable mode corresponding to the tangent to the NEB path as it passes over the barrier; this is a polar $B_u$ mode that reduces the symmetry from $C2/c$ to $Cc$. However, there is also a second $A_u$ mode that reduces the symmetry from $C2/c$ to $C2$. This latter antipolar mode has one Cu ion continuing on into the vdW gap, while the second Cu recedes back towards its original layer. The existence of this second unstable mode implies that $\overline{M}$ is not a plausible barrier structure and points in the direction of a new switching path featuring sequential Cu motion, in which the two Cu ions in the primitive cell traverse the vdW gaps one at a time. Therefore, our next step is to explore the possibility of such a switching path.

To explore such sequential paths, we distort the $\overline{M}$ structure along the $A_u$ mode and perform a full structural relaxation within $C2$; we obtain the structure we refer to as $R$, depicted in the middle panel of Fig.~\ref{fig:fig2}(b). The figure illustrates that unlike in the structures considered previously, where each Cu ion may be associated with a single layer, $R$ features Cu ions doubly occupying half of the layers, while leaving the other half empty. The 2-fold rotation axis of $R$ is found to lie along $x$ and to pass through the monolayer midplanes. We find that $R$ has no unstable zone center phonon modes, eliminating it as a barrier candidate.

To identify possible barriers, we perform an NEB path search initialized with the linearly interpolated path between $P^+$ and $R$, keeping the initial lattice vectors of each structure fixed during the calculation. However, we find that the calculation is difficult to converge because of the complexity of the path, which appears to pass over multiple saddle points on the way from $P^+$ to $R$.

To circumvent this difficulty, we resort to a two-stage procedure in which we first identify these saddle points in the energy landscape using a direct saddle-point search technique known as the dimer method \cite{dimer1-jcp1999,dimer2-jcp2005}. In this way we identify two saddle points which we refer to as $S^+_1$ and $S^+_2$, the latter of which is shown in the second panel of Fig.~\ref{fig:fig2}(b). These are then used as anchor points to split our NEB path into three segments $P^+\to S^+_1$, $S^+_1\to S^+_2$, and $S^+_2\to R$, and a separate NEB calculation for each segment is initialized with a linear interpolation of internal coordinates and lattice vectors between the respective endpoints. During the calculations, the initial lattice vectors of each structure on the NEB paths are held fixed.

The portion of the switching path from $R$ to $P^-$ is related to that from $P^+\to R$ by a $C_{2x}$ rotation symmetry. The structures with the highest energy are $S^{\pm}_1$, leading to an energy barrier of 0.91\,eV, much lower than the $\overline{M}$ barrier of 1.58\,eV reported above. Representative structures on the switching path are shown in Fig.~\ref{fig:fig2}(b). The Cu ions are found to pass through tetrahedral and coplanar trigonal local coordination environments, but never linear ones as was the case for the cooperative switching path shown in Fig.~\ref{fig:fig2}(a). Importantly, we emphasize that, unlike for the previous path, the Cu motion across the vdW gap is now sequential. That is, the Cu in the lower half of the primitive cell moves across the vdW gap first in Fig.~\ref{fig:fig2}(b), and once it has crossed, the second Cu traverses the upper vdW gap. 

The fact that the favored switching path through the vdW gaps is a sequential one raises the question of whether the same might be true of the ``ordinary'' switching path from $P^-$ to $P^+$, in which the Cu ions move within the monolayers~\cite{neal-prb2022}.  To investigate this, we use the PE phase structure - previously found to reside at the midpoint of the through-layer cooperative switching path - as a starting point in identifying a sequential switching path involving Cu motion through the layers. The PE structure features two unstable zone-center phonon modes: a polar $B_u$ mode that corresponds to the Cu off-centering instability driving the transition to the FE phase, and an antipolar $B_g$ mode that reduces the symmetry to $P\bar{1}$. We focus on the latter and relax the distorted structure while preserving the 20-atom translational symmetry of the PE phase. This results in a centrosymmetric structure, denoted as $I$, that is metastable (having no unstable zone-center modes) and features an energy of only 0.02\,eV relative to $P^-$ and $P^+$.

We perform an NEB calculation to find saddle points, which is initialized with structures residing on a linearly interpolated path between $I$ and $P^+$. In the course of the calculation, the initial lattice vectors of each structure are held fixed. The calculation reveals an energy barrier of 0.37\,eV, which is lower than that of the cooperative switching path through the layer. The path $P^-\to I$ is obtained by an inversion of the structures residing on $I\to P^+$.

Fig.~\ref{fig:fig2}(c) illustrates some of the representative structures on the switching path from $P^+$ to $P^-$ passing through $I$. As in the cross-gap switching path passing through $R$, the Cu ions move independently of each other. However, here they pass through the monolayers, with one Cu in the primitive cell initially passing through a layer followed by the second. 

\begin{figure}
\centering
\includegraphics[width= 8.5cm]{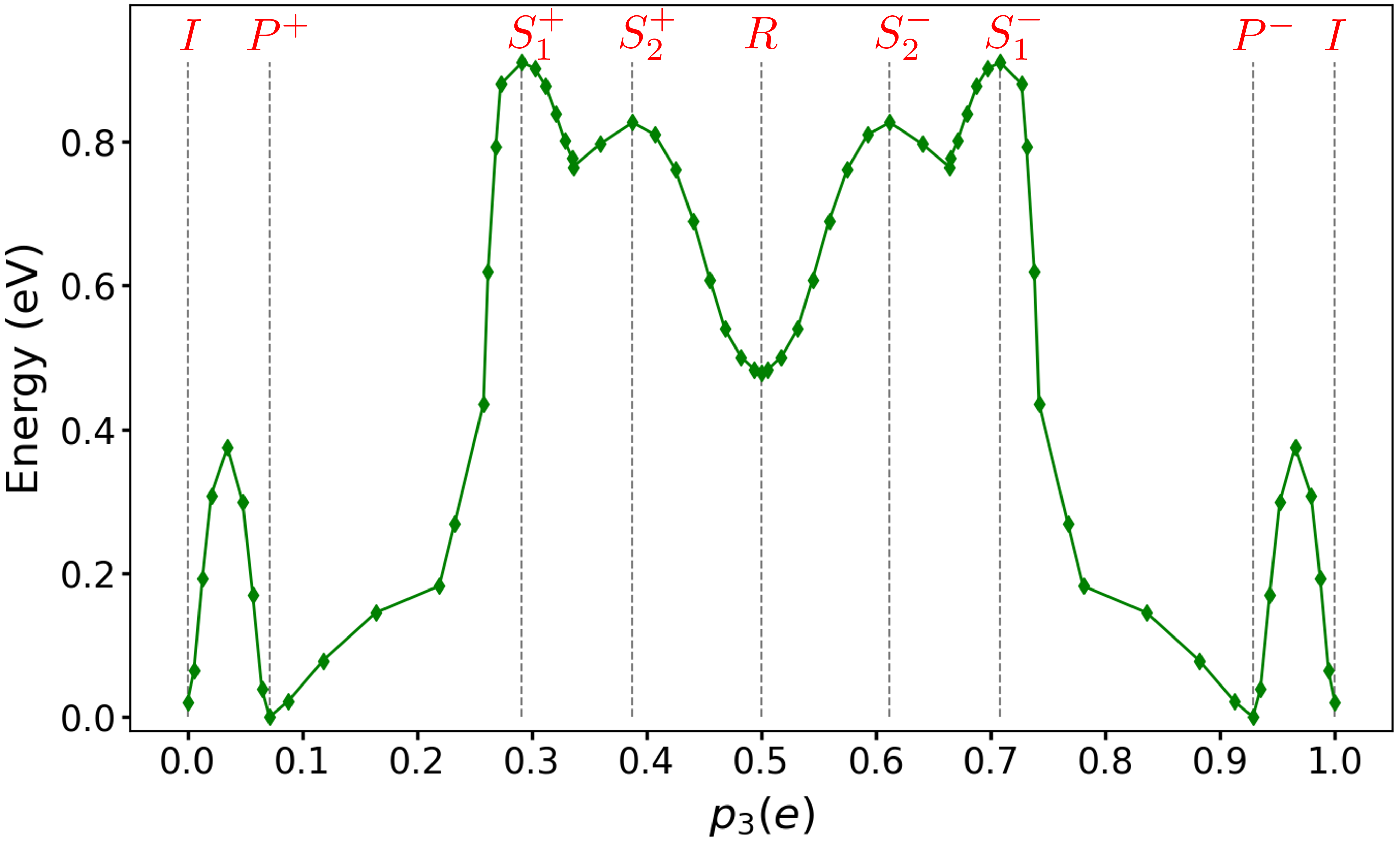}
\caption{The full NEB energy profile of the combined sequential switching paths, with Cu ions traversing the monolayers or vdW gaps, plotted versus the reduced polarization $p_3$. The most significant energy barriers for switching are defined by the difference in energies between $P^{\pm}$ and $S^{\pm}_1$ (cross-gap), and the local maxima on the $P^{\pm}\to I$ paths and $P^{\pm}$ (through-layer).}
\label{fig:fig3}
\end{figure}

The combination of the $P^+\to R\to P^-$ and $P^-\to I\to P^+$ switching paths described above yields a full cycle that brings $P^+$ back to itself, but with every Cu ion having moved vertically into the adjacent layer above. We would therefore like to understand the evolution of the OOP electric polarization along this cycle. This is only possible if the system remains insulating all along the path.  Crucially, we find that it does, with the smallest direct or indirect band gap of 0.88\,eV being found at the structures corresponding to $p_3=0.46$ and $0.54$ in Fig.~\ref{fig:fig3}. Since we are primarily interested in the polarization normal to the layers, we compute the reduced electric polarization $p_3=V_{\text{cell}}\textbf{b}_3\cdot\textbf{P}/2\pi$ for each configuration along the path.  Here $V_{\text{cell}}$ is the unit cell volume, $\textbf{b}_3$ is the third reciprocal lattice vector, and $\textbf{P}$ is the Berry-phase polarization. Note that $p_3$ is defined only modulo a quantum of elementary charge $e$.

Figure~\ref{fig:fig3} presents the total energy of each structure along the cycle on the vertical axis, and their corresponding reduced layer-normal polarizations $p_3$ on the horizontal axis. First, we note that the polarization remains well defined and increases monotonically along the cycle.  Moreover, we see that at the completion of the path, the polarization has changed by a quantum. The formal oxidation state of Cu in this compound is Cu$^{1+}$, consistent with the fact that one Cu ion from each primitive cell has moved into the vertically adjacent cell, thus realizing quantized charge transport \cite{jiang-prl2012}.

The presence of such a path suggests the possibility of adiabatic charge transport in this system.  An ideal scenario would be one in which a combination of two or more parameters, such as strain components or external fields, could be manipulated to carry the system deterministically along such a path, thereby realizing an adiabatic charge pump.  In the present case the barriers still appear to be too large to make this plausible, but further research is clearly desirable to see if more promising related systems can be found. 

Having specified our switching path, we would also like to obtain an order-of-magnitude estimate of the coercive fields involved in the switching. Ideally this could be done by using finite-field methods to follow the electric equation of state in the form of energy $E$ as a function of polarization $P_z$ normal to the layers~\cite{dieguez-prl2006,stengel-natphys2009,stengel-prb2009}, and associating the coercive field with the maximum of $dE/dP_z$ along the switching path. Since we are only interested in rough estimates here, we instead simply compute the finite difference $\Delta E/\Delta P_z$ on the most relevant segments of the switching path. Referring to Fig.~\ref{fig:fig3}, we find values of $0.77$, $0.24$, and $0.77$\,V/\AA\ in going from the minimum to the saddle point along $I$ towards $P^+$, along $R$ towards $S_2^-$, and along $P^-$ towards $I$, respectively. For the path from $P^+$ towards $S_1^+$ the finite difference is taken to start from the point at $p_3=0.26$, yielding a value of $1.21$\,V/\AA. It is noteworthy that this latter coercive field for switching through the vdW gap does not greatly exceed those required for switching through the layers.

It is well known that first-principles predictions of coercive fields based on preserving the primitive cell periodicity, such as those considered here, tend to overestimate experimentally measured values by a large amount, often by as much as two orders of magnitude. This is not surprising given that such calculations exclude the possibility of switching via domain-wall nucleation and motion and ignore the effects of temperature and disorder. In this context, our values, while high compared to corresponding values for conventional FEs, are not far out of line with experimental values for CIPS in the range of $10^{-4}$ to $10^{-3}$\,V/\r{A}~\cite{maisonneuve-prb1997,zhou-materhoriz2020}, which are also high compared with experimental coercive fields of conventional FEs.

To summarize, we investigated the energy landscape of CIPS to identify polarization switching paths involving either cooperative Cu ion motion preserving the system's glide mirror symmetry, or sequential Cu motion. Aside from the nature of the Cu motion, the two switching paths are distinguished by the local coordination environments that the Cu ions are found to visit in the course of the switching. The sequential switching paths through the monolayers and vdW gaps lead to significantly smaller energy barriers than their respective cooperative counterparts. The structures on the sequential switching paths are found to be insulating; we therefore argue that the combination of the two paths leads to a FE switching cycle that embodies the physics of a quantized adiabatic charge pump. Finally, we perform an order-of-magnitude estimate of the coercive fields needed to drive the system across the most significant energy barriers along the sequential switching cycle, and find that they are comparable in magnitude. 

Our findings reinforce an emerging focus on a class of materials that simultaneously exhibit both ferroelectricity and ionic conductivity in the bulk. In most systems, ionic transport is associated with the degradation of the FE state, and relatively few systems are known in which the two coexist \cite{habbal-jcp1978,scott-jcp1980,hoshino-ssi1981,stefanovich-FE1984,stefanovich-jjap1985,baranov-FE1989,scott-ssi1999}.
Our study offers an indication that this class of materials has the potential to exhibit unusually rich behavior.  Much remains to be understood, and further research is needed, regarding the relationship of ferroelectricity and ionic conductivity in these intriguing materials \cite{zhou-materhoriz2020,neumayer-ami2022,lindgren-ami2018,maglione-ome2016}.

\begin{acknowledgments}
DS and DV acknowledge support from the National Science Foundation under grants DGE-1842213 and DMR-1954856, respectively. KMR and SS acknowledge support from Office of Naval Research Grant N00014-21-1-2107, and JB acknowledges support from the Simons Foundation.
\end{acknowledgments}

\bibliography{bibfile.bib}

\begin{thebibliography}{46}%
\makeatletter
\providecommand \@ifxundefined [1]{%
 \@ifx{#1\undefined}
}%
\providecommand \@ifnum [1]{%
 \ifnum #1\expandafter \@firstoftwo
 \else \expandafter \@secondoftwo
 \fi
}%
\providecommand \@ifx [1]{%
 \ifx #1\expandafter \@firstoftwo
 \else \expandafter \@secondoftwo
 \fi
}%
\providecommand \natexlab [1]{#1}%
\providecommand \enquote  [1]{``#1''}%
\providecommand \bibnamefont  [1]{#1}%
\providecommand \bibfnamefont [1]{#1}%
\providecommand \citenamefont [1]{#1}%
\providecommand \href@noop [0]{\@secondoftwo}%
\providecommand \href [0]{\begingroup \@sanitize@url \@href}%
\providecommand \@href[1]{\@@startlink{#1}\@@href}%
\providecommand \@@href[1]{\endgroup#1\@@endlink}%
\providecommand \@sanitize@url [0]{\catcode `\\12\catcode `\$12\catcode
  `\&12\catcode `\#12\catcode `\^12\catcode `\_12\catcode `\%12\relax}%
\providecommand \@@startlink[1]{}%
\providecommand \@@endlink[0]{}%
\providecommand \url  [0]{\begingroup\@sanitize@url \@url }%
\providecommand \@url [1]{\endgroup\@href {#1}{\urlprefix }}%
\providecommand \urlprefix  [0]{URL }%
\providecommand \Eprint [0]{\href }%
\providecommand \doibase [0]{http://dx.doi.org/}%
\providecommand \selectlanguage [0]{\@gobble}%
\providecommand \bibinfo  [0]{\@secondoftwo}%
\providecommand \bibfield  [0]{\@secondoftwo}%
\providecommand \translation [1]{[#1]}%
\providecommand \BibitemOpen [0]{}%
\providecommand \bibitemStop [0]{}%
\providecommand \bibitemNoStop [0]{.\EOS\space}%
\providecommand \EOS [0]{\spacefactor3000\relax}%
\providecommand \BibitemShut  [1]{\csname bibitem#1\endcsname}%
\let\auto@bib@innerbib\@empty
\bibitem [{\citenamefont {Xue}\ \emph {et~al.}(2021)\citenamefont {Xue},
  \citenamefont {He},\ and\ \citenamefont {Zhang}}]{xue-apr2020}%
  \BibitemOpen
  \bibfield  {author} {\bibinfo {author} {\bibfnamefont {Fei}\ \bibnamefont
  {Xue}}, \bibinfo {author} {\bibfnamefont {Jr-Hau}\ \bibnamefont {He}}, \ and\
  \bibinfo {author} {\bibfnamefont {Xixiang}\ \bibnamefont {Zhang}},\
  }\bibfield  {title} {\enquote {\bibinfo {title} {Emerging van der {Waals}
  ferroelectrics: Unique properties and novel devices},}\ }\href {\doibase
  10.1063/5.0028079} {\bibfield  {journal} {\bibinfo  {journal} {Applied
  Physics Reviews}\ }\textbf {\bibinfo {volume} {8}},\ \bibinfo {pages}
  {021316} (\bibinfo {year} {2021})}\BibitemShut {NoStop}%
\bibitem [{\citenamefont {Zhang}\ \emph {et~al.}(2023)\citenamefont {Zhang},
  \citenamefont {Schoenherr}, \citenamefont {Sharma},\ and\ \citenamefont
  {Seidel}}]{zhang-natrevmat2022}%
  \BibitemOpen
  \bibfield  {author} {\bibinfo {author} {\bibfnamefont {Dawei}\ \bibnamefont
  {Zhang}}, \bibinfo {author} {\bibfnamefont {Peggy}\ \bibnamefont
  {Schoenherr}}, \bibinfo {author} {\bibfnamefont {Pankaj}\ \bibnamefont
  {Sharma}}, \ and\ \bibinfo {author} {\bibfnamefont {Jan}\ \bibnamefont
  {Seidel}},\ }\bibfield  {title} {\enquote {\bibinfo {title} {Ferroelectric
  order in van der {Waals} layered materials},}\ }\href {\doibase
  10.1038/s41578-022-00484-3} {\bibfield  {journal} {\bibinfo  {journal}
  {Nature Reviews Materials}\ }\textbf {\bibinfo {volume} {8}},\ \bibinfo
  {pages} {25--40} (\bibinfo {year} {2023})}\BibitemShut {NoStop}%
\bibitem [{\citenamefont {Zhou}\ \emph
  {et~al.}(2020{\natexlab{a}})\citenamefont {Zhou}, \citenamefont {You},
  \citenamefont {Zhou}, \citenamefont {Pu}, \citenamefont {Gui},\ and\
  \citenamefont {Wang}}]{zhou-fop2020}%
  \BibitemOpen
  \bibfield  {author} {\bibinfo {author} {\bibfnamefont {Shuang}\ \bibnamefont
  {Zhou}}, \bibinfo {author} {\bibfnamefont {Lu}~\bibnamefont {You}}, \bibinfo
  {author} {\bibfnamefont {Hailin}\ \bibnamefont {Zhou}}, \bibinfo {author}
  {\bibfnamefont {Yong}\ \bibnamefont {Pu}}, \bibinfo {author} {\bibfnamefont
  {Zhigang}\ \bibnamefont {Gui}}, \ and\ \bibinfo {author} {\bibfnamefont
  {Junling}\ \bibnamefont {Wang}},\ }\bibfield  {title} {\enquote {\bibinfo
  {title} {{Van der Waals} layered ferroelectric {CuInP$_2$S$_6$}: {Physical}
  properties and device applications},}\ }\href {\doibase
  10.1007/s11467-020-0986-0} {\bibfield  {journal} {\bibinfo  {journal}
  {Frontiers of Physics}\ }\textbf {\bibinfo {volume} {16}},\ \bibinfo {pages}
  {13301} (\bibinfo {year} {2020}{\natexlab{a}})}\BibitemShut {NoStop}%
\bibitem [{\citenamefont {Liu}\ \emph {et~al.}(2016)\citenamefont {Liu},
  \citenamefont {You}, \citenamefont {Seyler}, \citenamefont {Li},
  \citenamefont {Yu}, \citenamefont {Lin}, \citenamefont {Wang}, \citenamefont
  {Zhou}, \citenamefont {Wang}, \citenamefont {He}, \citenamefont {Pantelides},
  \citenamefont {Zhou}, \citenamefont {Sharma}, \citenamefont {Xu},
  \citenamefont {Ajayan}, \citenamefont {Wang},\ and\ \citenamefont
  {Liu}}]{liu-natcomm2016}%
  \BibitemOpen
  \bibfield  {author} {\bibinfo {author} {\bibfnamefont {Fucai}\ \bibnamefont
  {Liu}}, \bibinfo {author} {\bibfnamefont {Lu}~\bibnamefont {You}}, \bibinfo
  {author} {\bibfnamefont {Kyle~L.}\ \bibnamefont {Seyler}}, \bibinfo {author}
  {\bibfnamefont {Xiaobao}\ \bibnamefont {Li}}, \bibinfo {author}
  {\bibfnamefont {Peng}\ \bibnamefont {Yu}}, \bibinfo {author} {\bibfnamefont
  {Junhao}\ \bibnamefont {Lin}}, \bibinfo {author} {\bibfnamefont {Xuewen}\
  \bibnamefont {Wang}}, \bibinfo {author} {\bibfnamefont {Jiadong}\
  \bibnamefont {Zhou}}, \bibinfo {author} {\bibfnamefont {Hong}\ \bibnamefont
  {Wang}}, \bibinfo {author} {\bibfnamefont {Haiyong}\ \bibnamefont {He}},
  \bibinfo {author} {\bibfnamefont {Sokrates~T.}\ \bibnamefont {Pantelides}},
  \bibinfo {author} {\bibfnamefont {Wu}~\bibnamefont {Zhou}}, \bibinfo {author}
  {\bibfnamefont {Pradeep}\ \bibnamefont {Sharma}}, \bibinfo {author}
  {\bibfnamefont {Xiaodong}\ \bibnamefont {Xu}}, \bibinfo {author}
  {\bibfnamefont {Pulickel~M.}\ \bibnamefont {Ajayan}}, \bibinfo {author}
  {\bibfnamefont {Junling}\ \bibnamefont {Wang}}, \ and\ \bibinfo {author}
  {\bibfnamefont {Zheng}\ \bibnamefont {Liu}},\ }\bibfield  {title} {\enquote
  {\bibinfo {title} {Room-temperature ferroelectricity in {CuInP$_2$S$_6$}
  ultrathin flakes},}\ }\href {\doibase 10.1038/ncomms12357} {\bibfield
  {journal} {\bibinfo  {journal} {Nature Communications}\ }\textbf {\bibinfo
  {volume} {7}},\ \bibinfo {pages} {12357} (\bibinfo {year}
  {2016})}\BibitemShut {NoStop}%
\bibitem [{\citenamefont {Choi}\ \emph {et~al.}(2004)\citenamefont {Choi},
  \citenamefont {Biegalski}, \citenamefont {Li}, \citenamefont {Sharan},
  \citenamefont {Schubert}, \citenamefont {Uecker}, \citenamefont {Reiche},
  \citenamefont {Chen}, \citenamefont {Pan}, \citenamefont {Gopalan},
  \citenamefont {Chen}, \citenamefont {Schlom},\ and\ \citenamefont
  {Eom}}]{choi-science2004}%
  \BibitemOpen
  \bibfield  {author} {\bibinfo {author} {\bibfnamefont {K.~J.}\ \bibnamefont
  {Choi}}, \bibinfo {author} {\bibfnamefont {M.}~\bibnamefont {Biegalski}},
  \bibinfo {author} {\bibfnamefont {Y.~L.}\ \bibnamefont {Li}}, \bibinfo
  {author} {\bibfnamefont {A.}~\bibnamefont {Sharan}}, \bibinfo {author}
  {\bibfnamefont {J.}~\bibnamefont {Schubert}}, \bibinfo {author}
  {\bibfnamefont {R.}~\bibnamefont {Uecker}}, \bibinfo {author} {\bibfnamefont
  {P.}~\bibnamefont {Reiche}}, \bibinfo {author} {\bibfnamefont {Y.~B.}\
  \bibnamefont {Chen}}, \bibinfo {author} {\bibfnamefont {X.~Q.}\ \bibnamefont
  {Pan}}, \bibinfo {author} {\bibfnamefont {V.}~\bibnamefont {Gopalan}},
  \bibinfo {author} {\bibfnamefont {L.-Q.}\ \bibnamefont {Chen}}, \bibinfo
  {author} {\bibfnamefont {D.~G.}\ \bibnamefont {Schlom}}, \ and\ \bibinfo
  {author} {\bibfnamefont {C.~B.}\ \bibnamefont {Eom}},\ }\bibfield  {title}
  {\enquote {\bibinfo {title} {Enhancement of ferroelectricity in strained
  {BaTiO$_3$} thin films},}\ }\href {\doibase 10.1126/science.1103218}
  {\bibfield  {journal} {\bibinfo  {journal} {Science}\ }\textbf {\bibinfo
  {volume} {306}},\ \bibinfo {pages} {1005--1009} (\bibinfo {year}
  {2004})}\BibitemShut {NoStop}%
\bibitem [{\citenamefont {Garcia}\ \emph {et~al.}(2009)\citenamefont {Garcia},
  \citenamefont {Fusil}, \citenamefont {Bouzehouane}, \citenamefont
  {Enouz-Vedrenne}, \citenamefont {Mathur}, \citenamefont
  {Barth{\'e}l{\'e}my},\ and\ \citenamefont {Bibes}}]{garcia-nature2009}%
  \BibitemOpen
  \bibfield  {author} {\bibinfo {author} {\bibfnamefont {V.}~\bibnamefont
  {Garcia}}, \bibinfo {author} {\bibfnamefont {S.}~\bibnamefont {Fusil}},
  \bibinfo {author} {\bibfnamefont {K.}~\bibnamefont {Bouzehouane}}, \bibinfo
  {author} {\bibfnamefont {S.}~\bibnamefont {Enouz-Vedrenne}}, \bibinfo
  {author} {\bibfnamefont {N.~D.}\ \bibnamefont {Mathur}}, \bibinfo {author}
  {\bibfnamefont {A.}~\bibnamefont {Barth{\'e}l{\'e}my}}, \ and\ \bibinfo
  {author} {\bibfnamefont {M.}~\bibnamefont {Bibes}},\ }\bibfield  {title}
  {\enquote {\bibinfo {title} {Giant tunnel electroresistance for
  non-destructive readout of ferroelectric states},}\ }\href {\doibase
  10.1038/nature08128} {\bibfield  {journal} {\bibinfo  {journal} {Nature}\
  }\textbf {\bibinfo {volume} {460}},\ \bibinfo {pages} {81--84} (\bibinfo
  {year} {2009})}\BibitemShut {NoStop}%
\bibitem [{\citenamefont {Zhang}\ \emph {et~al.}(2014)\citenamefont {Zhang},
  \citenamefont {Li}, \citenamefont {Shimada}, \citenamefont {Wang},\ and\
  \citenamefont {Kitamura}}]{zhang-prb2014}%
  \BibitemOpen
  \bibfield  {author} {\bibinfo {author} {\bibfnamefont {Yajun}\ \bibnamefont
  {Zhang}}, \bibinfo {author} {\bibfnamefont {Gui-Ping}\ \bibnamefont {Li}},
  \bibinfo {author} {\bibfnamefont {Takahiro}\ \bibnamefont {Shimada}},
  \bibinfo {author} {\bibfnamefont {Jie}\ \bibnamefont {Wang}}, \ and\ \bibinfo
  {author} {\bibfnamefont {Takayuki}\ \bibnamefont {Kitamura}},\ }\bibfield
  {title} {\enquote {\bibinfo {title} {Disappearance of ferroelectric critical
  thickness in epitaxial ultrathin {BaZrO$_3$} films},}\ }\href {\doibase
  10.1103/PhysRevB.90.184107} {\bibfield  {journal} {\bibinfo  {journal} {Phys.
  Rev. B}\ }\textbf {\bibinfo {volume} {90}},\ \bibinfo {pages} {184107}
  (\bibinfo {year} {2014})}\BibitemShut {NoStop}%
\bibitem [{\citenamefont {Sai}\ \emph {et~al.}(2005)\citenamefont {Sai},
  \citenamefont {Kolpak},\ and\ \citenamefont {Rappe}}]{sai-prb2005}%
  \BibitemOpen
  \bibfield  {author} {\bibinfo {author} {\bibfnamefont {Na}~\bibnamefont
  {Sai}}, \bibinfo {author} {\bibfnamefont {Alexie~M.}\ \bibnamefont {Kolpak}},
  \ and\ \bibinfo {author} {\bibfnamefont {Andrew~M.}\ \bibnamefont {Rappe}},\
  }\bibfield  {title} {\enquote {\bibinfo {title} {Ferroelectricity in
  ultrathin perovskite films},}\ }\href {\doibase 10.1103/PhysRevB.72.020101}
  {\bibfield  {journal} {\bibinfo  {journal} {Phys. Rev. B}\ }\textbf {\bibinfo
  {volume} {72}},\ \bibinfo {pages} {020101} (\bibinfo {year}
  {2005})}\BibitemShut {NoStop}%
\bibitem [{\citenamefont {You}\ \emph {et~al.}(2019)\citenamefont {You},
  \citenamefont {Zhang}, \citenamefont {Zhou}, \citenamefont {Chaturvedi},
  \citenamefont {Morris}, \citenamefont {Liu}, \citenamefont {Chang},
  \citenamefont {Ichinose}, \citenamefont {Funakubo}, \citenamefont {Hu},
  \citenamefont {Wu}, \citenamefont {Liu}, \citenamefont {Dong},\ and\
  \citenamefont {Wang}}]{you-sciadv2019}%
  \BibitemOpen
  \bibfield  {author} {\bibinfo {author} {\bibfnamefont {Lu}~\bibnamefont
  {You}}, \bibinfo {author} {\bibfnamefont {Yang}\ \bibnamefont {Zhang}},
  \bibinfo {author} {\bibfnamefont {Shuang}\ \bibnamefont {Zhou}}, \bibinfo
  {author} {\bibfnamefont {Apoorva}\ \bibnamefont {Chaturvedi}}, \bibinfo
  {author} {\bibfnamefont {Samuel~A.}\ \bibnamefont {Morris}}, \bibinfo
  {author} {\bibfnamefont {Fucai}\ \bibnamefont {Liu}}, \bibinfo {author}
  {\bibfnamefont {Lei}\ \bibnamefont {Chang}}, \bibinfo {author} {\bibfnamefont
  {Daichi}\ \bibnamefont {Ichinose}}, \bibinfo {author} {\bibfnamefont
  {Hiroshi}\ \bibnamefont {Funakubo}}, \bibinfo {author} {\bibfnamefont
  {Weijin}\ \bibnamefont {Hu}}, \bibinfo {author} {\bibfnamefont {Tom}\
  \bibnamefont {Wu}}, \bibinfo {author} {\bibfnamefont {Zheng}\ \bibnamefont
  {Liu}}, \bibinfo {author} {\bibfnamefont {Shuai}\ \bibnamefont {Dong}}, \
  and\ \bibinfo {author} {\bibfnamefont {Junling}\ \bibnamefont {Wang}},\
  }\bibfield  {title} {\enquote {\bibinfo {title} {Origin of giant negative
  piezoelectricity in a layered van der {Waals} ferroelectric},}\ }\href
  {\doibase 10.1126/sciadv.aav3780} {\bibfield  {journal} {\bibinfo  {journal}
  {Science Advances}\ }\textbf {\bibinfo {volume} {5}},\ \bibinfo {pages}
  {eaav3780} (\bibinfo {year} {2019})}\BibitemShut {NoStop}%
\bibitem [{\citenamefont {Brehm}\ \emph {et~al.}(2020)\citenamefont {Brehm},
  \citenamefont {Neumayer}, \citenamefont {Tao}, \citenamefont {O'Hara},
  \citenamefont {Chyasnavichus}, \citenamefont {Susner}, \citenamefont
  {McGuire}, \citenamefont {Kalinin}, \citenamefont {Jesse}, \citenamefont
  {Ganesh}, \citenamefont {Pantelides}, \citenamefont {Maksymovych},\ and\
  \citenamefont {Balke}}]{brehm-natmat2020}%
  \BibitemOpen
  \bibfield  {author} {\bibinfo {author} {\bibfnamefont {John~A.}\ \bibnamefont
  {Brehm}}, \bibinfo {author} {\bibfnamefont {Sabine~M.}\ \bibnamefont
  {Neumayer}}, \bibinfo {author} {\bibfnamefont {Lei}\ \bibnamefont {Tao}},
  \bibinfo {author} {\bibfnamefont {Andrew}\ \bibnamefont {O'Hara}}, \bibinfo
  {author} {\bibfnamefont {Marius}\ \bibnamefont {Chyasnavichus}}, \bibinfo
  {author} {\bibfnamefont {Michael~A.}\ \bibnamefont {Susner}}, \bibinfo
  {author} {\bibfnamefont {Michael~A.}\ \bibnamefont {McGuire}}, \bibinfo
  {author} {\bibfnamefont {Sergei~V.}\ \bibnamefont {Kalinin}}, \bibinfo
  {author} {\bibfnamefont {Stephen}\ \bibnamefont {Jesse}}, \bibinfo {author}
  {\bibfnamefont {Panchapakesan}\ \bibnamefont {Ganesh}}, \bibinfo {author}
  {\bibfnamefont {Sokrates~T.}\ \bibnamefont {Pantelides}}, \bibinfo {author}
  {\bibfnamefont {Petro}\ \bibnamefont {Maksymovych}}, \ and\ \bibinfo {author}
  {\bibfnamefont {Nina}\ \bibnamefont {Balke}},\ }\bibfield  {title} {\enquote
  {\bibinfo {title} {Tunable quadruple-well ferroelectric van der {Waals}
  crystals},}\ }\href {\doibase 10.1038/s41563-019-0532-z} {\bibfield
  {journal} {\bibinfo  {journal} {Nature Materials}\ }\textbf {\bibinfo
  {volume} {19}},\ \bibinfo {pages} {43--48} (\bibinfo {year}
  {2020})}\BibitemShut {NoStop}%
\bibitem [{\citenamefont {Neumayer}\ \emph {et~al.}(2019)\citenamefont
  {Neumayer}, \citenamefont {Eliseev}, \citenamefont {Susner}, \citenamefont
  {Tselev}, \citenamefont {Rodriguez}, \citenamefont {Brehm}, \citenamefont
  {Pantelides}, \citenamefont {Panchapakesan}, \citenamefont {Jesse},
  \citenamefont {Kalinin}, \citenamefont {McGuire}, \citenamefont {Morozovska},
  \citenamefont {Maksymovych},\ and\ \citenamefont {Balke}}]{neumayer-prm2019}%
  \BibitemOpen
  \bibfield  {author} {\bibinfo {author} {\bibfnamefont {Sabine~M.}\
  \bibnamefont {Neumayer}}, \bibinfo {author} {\bibfnamefont {Eugene~A.}\
  \bibnamefont {Eliseev}}, \bibinfo {author} {\bibfnamefont {Michael~A.}\
  \bibnamefont {Susner}}, \bibinfo {author} {\bibfnamefont {Alexander}\
  \bibnamefont {Tselev}}, \bibinfo {author} {\bibfnamefont {Brian~J.}\
  \bibnamefont {Rodriguez}}, \bibinfo {author} {\bibfnamefont {John~A.}\
  \bibnamefont {Brehm}}, \bibinfo {author} {\bibfnamefont {Sokrates~T.}\
  \bibnamefont {Pantelides}}, \bibinfo {author} {\bibfnamefont {Ganesh}\
  \bibnamefont {Panchapakesan}}, \bibinfo {author} {\bibfnamefont {Stephen}\
  \bibnamefont {Jesse}}, \bibinfo {author} {\bibfnamefont {Sergei~V.}\
  \bibnamefont {Kalinin}}, \bibinfo {author} {\bibfnamefont {Michael~A.}\
  \bibnamefont {McGuire}}, \bibinfo {author} {\bibfnamefont {Anna~N.}\
  \bibnamefont {Morozovska}}, \bibinfo {author} {\bibfnamefont {Petro}\
  \bibnamefont {Maksymovych}}, \ and\ \bibinfo {author} {\bibfnamefont {Nina}\
  \bibnamefont {Balke}},\ }\bibfield  {title} {\enquote {\bibinfo {title}
  {Giant negative electrostriction and dielectric tunability in a van der
  {Waals} layered ferroelectric},}\ }\href {\doibase
  10.1103/PhysRevMaterials.3.024401} {\bibfield  {journal} {\bibinfo  {journal}
  {Phys. Rev. Materials}\ }\textbf {\bibinfo {volume} {3}},\ \bibinfo {pages}
  {024401} (\bibinfo {year} {2019})}\BibitemShut {NoStop}%
\bibitem [{\citenamefont {Neumayer}\ \emph
  {et~al.}(2020{\natexlab{a}})\citenamefont {Neumayer}, \citenamefont {Tao},
  \citenamefont {O'Hara}, \citenamefont {Susner}, \citenamefont {McGuire},
  \citenamefont {Maksymovych}, \citenamefont {Pantelides},\ and\ \citenamefont
  {Balke}}]{neumayer-aem2020}%
  \BibitemOpen
  \bibfield  {author} {\bibinfo {author} {\bibfnamefont {Sabine~M.}\
  \bibnamefont {Neumayer}}, \bibinfo {author} {\bibfnamefont {Lei}\
  \bibnamefont {Tao}}, \bibinfo {author} {\bibfnamefont {Andrew}\ \bibnamefont
  {O'Hara}}, \bibinfo {author} {\bibfnamefont {Michael~A.}\ \bibnamefont
  {Susner}}, \bibinfo {author} {\bibfnamefont {Michael~A.}\ \bibnamefont
  {McGuire}}, \bibinfo {author} {\bibfnamefont {Petro}\ \bibnamefont
  {Maksymovych}}, \bibinfo {author} {\bibfnamefont {Sokrates~T.}\ \bibnamefont
  {Pantelides}}, \ and\ \bibinfo {author} {\bibfnamefont {Nina}\ \bibnamefont
  {Balke}},\ }\bibfield  {title} {\enquote {\bibinfo {title} {The concept of
  negative capacitance in ionically conductive van der {Waals}
  ferroelectrics},}\ }\href {\doibase https://doi.org/10.1002/aenm.202001726}
  {\bibfield  {journal} {\bibinfo  {journal} {Advanced Energy Materials}\
  }\textbf {\bibinfo {volume} {10}},\ \bibinfo {pages} {2001726} (\bibinfo
  {year} {2020}{\natexlab{a}})}\BibitemShut {NoStop}%
\bibitem [{\citenamefont {O'Hara}\ \emph {et~al.}(2022)\citenamefont {O'Hara},
  \citenamefont {Balke},\ and\ \citenamefont {Pantelides}}]{ohara-aem2022}%
  \BibitemOpen
  \bibfield  {author} {\bibinfo {author} {\bibfnamefont {Andrew}\ \bibnamefont
  {O'Hara}}, \bibinfo {author} {\bibfnamefont {Nina}\ \bibnamefont {Balke}}, \
  and\ \bibinfo {author} {\bibfnamefont {Sokrates~T.}\ \bibnamefont
  {Pantelides}},\ }\bibfield  {title} {\enquote {\bibinfo {title} {Unique
  features of polarization in ferroelectric ionic conductors},}\ }\href
  {\doibase https://doi.org/10.1002/aelm.202100810} {\bibfield  {journal}
  {\bibinfo  {journal} {Advanced Electronic Materials}\ }\textbf {\bibinfo
  {volume} {8}},\ \bibinfo {pages} {2100810} (\bibinfo {year}
  {2022})}\BibitemShut {NoStop}%
\bibitem [{\citenamefont {Maisonneuve}\ \emph
  {et~al.}(1997{\natexlab{a}})\citenamefont {Maisonneuve}, \citenamefont
  {Cajipe}, \citenamefont {Simon}, \citenamefont {Von Der~Muhll},\ and\
  \citenamefont {Ravez}}]{maisonneuve-prb1997}%
  \BibitemOpen
  \bibfield  {author} {\bibinfo {author} {\bibfnamefont {V.}~\bibnamefont
  {Maisonneuve}}, \bibinfo {author} {\bibfnamefont {V.~B.}\ \bibnamefont
  {Cajipe}}, \bibinfo {author} {\bibfnamefont {A.}~\bibnamefont {Simon}},
  \bibinfo {author} {\bibfnamefont {R.}~\bibnamefont {Von Der~Muhll}}, \ and\
  \bibinfo {author} {\bibfnamefont {J.}~\bibnamefont {Ravez}},\ }\bibfield
  {title} {\enquote {\bibinfo {title} {Ferrielectric ordering in lamellar
  {CuInP$_2$S$_6$}},}\ }\href {\doibase 10.1103/PhysRevB.56.10860} {\bibfield
  {journal} {\bibinfo  {journal} {Phys. Rev. B}\ }\textbf {\bibinfo {volume}
  {56}},\ \bibinfo {pages} {10860--10868} (\bibinfo {year}
  {1997}{\natexlab{a}})}\BibitemShut {NoStop}%
\bibitem [{\citenamefont {Vysochanskii}\ \emph {et~al.}(1998)\citenamefont
  {Vysochanskii}, \citenamefont {Stephanovich}, \citenamefont {Molnar},
  \citenamefont {Cajipe},\ and\ \citenamefont {Bourdon}}]{vysochanskii-1998}%
  \BibitemOpen
  \bibfield  {author} {\bibinfo {author} {\bibfnamefont {Yu.~M.}\ \bibnamefont
  {Vysochanskii}}, \bibinfo {author} {\bibfnamefont {V.~A.}\ \bibnamefont
  {Stephanovich}}, \bibinfo {author} {\bibfnamefont {A.~A.}\ \bibnamefont
  {Molnar}}, \bibinfo {author} {\bibfnamefont {V.~B.}\ \bibnamefont {Cajipe}},
  \ and\ \bibinfo {author} {\bibfnamefont {X.}~\bibnamefont {Bourdon}},\
  }\bibfield  {title} {\enquote {\bibinfo {title} {Raman spectroscopy study of
  the ferrielectric-paraelectric transition in layered {CuInP$_2$S$_6$}},}\
  }\href {\doibase 10.1103/PhysRevB.58.9119} {\bibfield  {journal} {\bibinfo
  {journal} {Phys. Rev. B}\ }\textbf {\bibinfo {volume} {58}},\ \bibinfo
  {pages} {9119--9124} (\bibinfo {year} {1998})}\BibitemShut {NoStop}%
\bibitem [{\citenamefont {Zhou}\ \emph
  {et~al.}(2020{\natexlab{b}})\citenamefont {Zhou}, \citenamefont {You},
  \citenamefont {Chaturvedi}, \citenamefont {Morris}, \citenamefont {Herrin},
  \citenamefont {Zhang}, \citenamefont {Abdelsamie}, \citenamefont {Hu},
  \citenamefont {Chen}, \citenamefont {Zhou}, \citenamefont {Dong},\ and\
  \citenamefont {Wang}}]{zhou-materhoriz2020}%
  \BibitemOpen
  \bibfield  {author} {\bibinfo {author} {\bibfnamefont {Shuang}\ \bibnamefont
  {Zhou}}, \bibinfo {author} {\bibfnamefont {Lu}~\bibnamefont {You}}, \bibinfo
  {author} {\bibfnamefont {Apoorva}\ \bibnamefont {Chaturvedi}}, \bibinfo
  {author} {\bibfnamefont {Samuel~A.}\ \bibnamefont {Morris}}, \bibinfo
  {author} {\bibfnamefont {Jason~S.}\ \bibnamefont {Herrin}}, \bibinfo {author}
  {\bibfnamefont {Na}~\bibnamefont {Zhang}}, \bibinfo {author} {\bibfnamefont
  {Amr}\ \bibnamefont {Abdelsamie}}, \bibinfo {author} {\bibfnamefont
  {Yuzhong}\ \bibnamefont {Hu}}, \bibinfo {author} {\bibfnamefont {Jieqiong}\
  \bibnamefont {Chen}}, \bibinfo {author} {\bibfnamefont {Yang}\ \bibnamefont
  {Zhou}}, \bibinfo {author} {\bibfnamefont {Shuai}\ \bibnamefont {Dong}}, \
  and\ \bibinfo {author} {\bibfnamefont {Junling}\ \bibnamefont {Wang}},\
  }\bibfield  {title} {\enquote {\bibinfo {title} {Anomalous polarization
  switching and permanent retention in a ferroelectric ionic conductor},}\
  }\href {\doibase 10.1039/C9MH01215J} {\bibfield  {journal} {\bibinfo
  {journal} {Mater. Horiz.}\ }\textbf {\bibinfo {volume} {7}},\ \bibinfo
  {pages} {263--274} (\bibinfo {year} {2020}{\natexlab{b}})}\BibitemShut
  {NoStop}%
\bibitem [{\citenamefont {Balke}\ \emph {et~al.}(2018)\citenamefont {Balke},
  \citenamefont {Neumayer}, \citenamefont {Brehm}, \citenamefont {Susner},
  \citenamefont {Rodriguez}, \citenamefont {Jesse}, \citenamefont {Kalinin},
  \citenamefont {Pantelides}, \citenamefont {McGuire},\ and\ \citenamefont
  {Maksymovych}}]{balke-ami2018}%
  \BibitemOpen
  \bibfield  {author} {\bibinfo {author} {\bibfnamefont {Nina}\ \bibnamefont
  {Balke}}, \bibinfo {author} {\bibfnamefont {Sabine~M.}\ \bibnamefont
  {Neumayer}}, \bibinfo {author} {\bibfnamefont {John~A.}\ \bibnamefont
  {Brehm}}, \bibinfo {author} {\bibfnamefont {Michael~A.}\ \bibnamefont
  {Susner}}, \bibinfo {author} {\bibfnamefont {Brian~J.}\ \bibnamefont
  {Rodriguez}}, \bibinfo {author} {\bibfnamefont {Stephen}\ \bibnamefont
  {Jesse}}, \bibinfo {author} {\bibfnamefont {Sergei~V.}\ \bibnamefont
  {Kalinin}}, \bibinfo {author} {\bibfnamefont {Sokrates~T.}\ \bibnamefont
  {Pantelides}}, \bibinfo {author} {\bibfnamefont {Michael~A.}\ \bibnamefont
  {McGuire}}, \ and\ \bibinfo {author} {\bibfnamefont {Petro}\ \bibnamefont
  {Maksymovych}},\ }\bibfield  {title} {\enquote {\bibinfo {title} {Locally
  controlled {Cu}-ion transport in layered ferroelectric {CuInP$_2$S$_6$}},}\
  }\href {\doibase 10.1021/acsami.8b08079} {\bibfield  {journal} {\bibinfo
  {journal} {ACS Applied Materials \& Interfaces}\ }\textbf {\bibinfo {volume}
  {10}},\ \bibinfo {pages} {27188--27194} (\bibinfo {year} {2018})}\BibitemShut
  {NoStop}%
\bibitem [{\citenamefont {Maisonneuve}\ \emph
  {et~al.}(1997{\natexlab{b}})\citenamefont {Maisonneuve}, \citenamefont
  {Reau}, \citenamefont {Dong}, \citenamefont {Cajipe}, \citenamefont {Payen},\
  and\ \citenamefont {Ravez}}]{maisonneuve-FE1997}%
  \BibitemOpen
  \bibfield  {author} {\bibinfo {author} {\bibfnamefont {V.}~\bibnamefont
  {Maisonneuve}}, \bibinfo {author} {\bibfnamefont {J.~M.}\ \bibnamefont
  {Reau}}, \bibinfo {author} {\bibfnamefont {Ming}\ \bibnamefont {Dong}},
  \bibinfo {author} {\bibfnamefont {V.~B.}\ \bibnamefont {Cajipe}}, \bibinfo
  {author} {\bibfnamefont {C.}~\bibnamefont {Payen}}, \ and\ \bibinfo {author}
  {\bibfnamefont {J.}~\bibnamefont {Ravez}},\ }\bibfield  {title} {\enquote
  {\bibinfo {title} {Ionic conductivity in ferroic {CuInP$_2$S$_6$} and
  {CuCrP$_2$S$_6$}},}\ }\href {\doibase 10.1080/00150199708224175} {\bibfield
  {journal} {\bibinfo  {journal} {Ferroelectrics}\ }\textbf {\bibinfo {volume}
  {196}},\ \bibinfo {pages} {257--260} (\bibinfo {year}
  {1997}{\natexlab{b}})}\BibitemShut {NoStop}%
\bibitem [{\citenamefont {Banys}\ \emph {et~al.}(2004)\citenamefont {Banys},
  \citenamefont {Macutkevic}, \citenamefont {Samulionis}, \citenamefont
  {Brilingas},\ and\ \citenamefont {Vysochanskii}}]{banys-pt2004}%
  \BibitemOpen
  \bibfield  {author} {\bibinfo {author} {\bibfnamefont {J.}~\bibnamefont
  {Banys}}, \bibinfo {author} {\bibfnamefont {J.}~\bibnamefont {Macutkevic}},
  \bibinfo {author} {\bibfnamefont {V.}~\bibnamefont {Samulionis}}, \bibinfo
  {author} {\bibfnamefont {A.}~\bibnamefont {Brilingas}}, \ and\ \bibinfo
  {author} {\bibfnamefont {Yu.}\ \bibnamefont {Vysochanskii}},\ }\bibfield
  {title} {\enquote {\bibinfo {title} {Dielectric and ultrasonic investigation
  of phase transition in {CuInP$_2$S$_6$} crystals},}\ }\href {\doibase
  10.1080/01411590410001667608} {\bibfield  {journal} {\bibinfo  {journal}
  {Phase Transitions}\ }\textbf {\bibinfo {volume} {77}},\ \bibinfo {pages}
  {345--358} (\bibinfo {year} {2004})}\BibitemShut {NoStop}%
\bibitem [{\citenamefont {Dziaugys}\ \emph {et~al.}(2013)\citenamefont
  {Dziaugys}, \citenamefont {Banys}, \citenamefont {Macutkevic},\ and\
  \citenamefont {Vysochanskii}}]{dziaugys-pt2013}%
  \BibitemOpen
  \bibfield  {author} {\bibinfo {author} {\bibfnamefont {Andrius}\ \bibnamefont
  {Dziaugys}}, \bibinfo {author} {\bibfnamefont {Juras}\ \bibnamefont {Banys}},
  \bibinfo {author} {\bibfnamefont {Jan}\ \bibnamefont {Macutkevic}}, \ and\
  \bibinfo {author} {\bibfnamefont {Yulian}\ \bibnamefont {Vysochanskii}},\
  }\bibfield  {title} {\enquote {\bibinfo {title} {Anisotropy effects in thick
  layered {CuInP$_2$S$_6$} and {CuInP$_2$Se$_6$} crystals},}\ }\href {\doibase
  10.1080/01411594.2012.745533} {\bibfield  {journal} {\bibinfo  {journal}
  {Phase Transitions}\ }\textbf {\bibinfo {volume} {86}},\ \bibinfo {pages}
  {878--885} (\bibinfo {year} {2013})}\BibitemShut {NoStop}%
\bibitem [{\citenamefont {Macutkevic}\ \emph {et~al.}(2015)\citenamefont
  {Macutkevic}, \citenamefont {Banys},\ and\ \citenamefont
  {Vysochanskii}}]{macutkevic-pssb2015}%
  \BibitemOpen
  \bibfield  {author} {\bibinfo {author} {\bibfnamefont {Jan}\ \bibnamefont
  {Macutkevic}}, \bibinfo {author} {\bibfnamefont {Juras}\ \bibnamefont
  {Banys}}, \ and\ \bibinfo {author} {\bibfnamefont {Yulian}\ \bibnamefont
  {Vysochanskii}},\ }\bibfield  {title} {\enquote {\bibinfo {title} {Electrical
  conductivity of layered {CuInP$_2$(S$_x$Se$_{1-x}$)$_6$} crystals},}\ }\href
  {\doibase 10.1002/pssb.201451738} {\bibfield  {journal} {\bibinfo  {journal}
  {physica status solidi (b)}\ }\textbf {\bibinfo {volume} {252}},\ \bibinfo
  {pages} {1773--1777} (\bibinfo {year} {2015})}\BibitemShut {NoStop}%
\bibitem [{\citenamefont {Neal}\ \emph {et~al.}(2022)\citenamefont {Neal},
  \citenamefont {Singh}, \citenamefont {Fang}, \citenamefont {Won},
  \citenamefont {Huang}, \citenamefont {Cheong}, \citenamefont {Rabe},
  \citenamefont {Vanderbilt},\ and\ \citenamefont {Musfeldt}}]{neal-prb2022}%
  \BibitemOpen
  \bibfield  {author} {\bibinfo {author} {\bibfnamefont {Sabine~N.}\
  \bibnamefont {Neal}}, \bibinfo {author} {\bibfnamefont {Sobhit}\ \bibnamefont
  {Singh}}, \bibinfo {author} {\bibfnamefont {Xiaochen}\ \bibnamefont {Fang}},
  \bibinfo {author} {\bibfnamefont {Choongjae}\ \bibnamefont {Won}}, \bibinfo
  {author} {\bibfnamefont {Fei-Ting}\ \bibnamefont {Huang}}, \bibinfo {author}
  {\bibfnamefont {Sang-Wook}\ \bibnamefont {Cheong}}, \bibinfo {author}
  {\bibfnamefont {Karin~M.}\ \bibnamefont {Rabe}}, \bibinfo {author}
  {\bibfnamefont {David}\ \bibnamefont {Vanderbilt}}, \ and\ \bibinfo {author}
  {\bibfnamefont {Janice~L.}\ \bibnamefont {Musfeldt}},\ }\bibfield  {title}
  {\enquote {\bibinfo {title} {Vibrational properties of {CuInP$_2$S$_6$}
  across the ferroelectric transition},}\ }\href {\doibase
  10.1103/PhysRevB.105.075151} {\bibfield  {journal} {\bibinfo  {journal}
  {Phys. Rev. B}\ }\textbf {\bibinfo {volume} {105}},\ \bibinfo {pages}
  {075151} (\bibinfo {year} {2022})}\BibitemShut {NoStop}%
\bibitem [{\citenamefont {Neumayer}\ \emph
  {et~al.}(2020{\natexlab{b}})\citenamefont {Neumayer}, \citenamefont {Tao},
  \citenamefont {O'Hara}, \citenamefont {Brehm}, \citenamefont {Si},
  \citenamefont {Liao}, \citenamefont {Feng}, \citenamefont {Kalinin},
  \citenamefont {Ye}, \citenamefont {Pantelides}, \citenamefont {Maksymovych},\
  and\ \citenamefont {Balke}}]{neumayer-pra2020}%
  \BibitemOpen
  \bibfield  {author} {\bibinfo {author} {\bibfnamefont {Sabine~M.}\
  \bibnamefont {Neumayer}}, \bibinfo {author} {\bibfnamefont {Lei}\
  \bibnamefont {Tao}}, \bibinfo {author} {\bibfnamefont {Andrew}\ \bibnamefont
  {O'Hara}}, \bibinfo {author} {\bibfnamefont {John}\ \bibnamefont {Brehm}},
  \bibinfo {author} {\bibfnamefont {Mengwei}\ \bibnamefont {Si}}, \bibinfo
  {author} {\bibfnamefont {Pai-Ying}\ \bibnamefont {Liao}}, \bibinfo {author}
  {\bibfnamefont {Tianli}\ \bibnamefont {Feng}}, \bibinfo {author}
  {\bibfnamefont {Sergei~V.}\ \bibnamefont {Kalinin}}, \bibinfo {author}
  {\bibfnamefont {Peide~D.}\ \bibnamefont {Ye}}, \bibinfo {author}
  {\bibfnamefont {Sokrates~T.}\ \bibnamefont {Pantelides}}, \bibinfo {author}
  {\bibfnamefont {Petro}\ \bibnamefont {Maksymovych}}, \ and\ \bibinfo {author}
  {\bibfnamefont {Nina}\ \bibnamefont {Balke}},\ }\bibfield  {title} {\enquote
  {\bibinfo {title} {Alignment of polarization against an electric field in van
  der {Waals} ferroelectrics},}\ }\href {\doibase
  10.1103/PhysRevApplied.13.064063} {\bibfield  {journal} {\bibinfo  {journal}
  {Phys. Rev. Applied}\ }\textbf {\bibinfo {volume} {13}},\ \bibinfo {pages}
  {064063} (\bibinfo {year} {2020}{\natexlab{b}})}\BibitemShut {NoStop}%
\bibitem [{\citenamefont {Zhang}\ \emph {et~al.}(2021)\citenamefont {Zhang},
  \citenamefont {Luo}, \citenamefont {Yao}, \citenamefont {Schoenherr},
  \citenamefont {Sha}, \citenamefont {Pan}, \citenamefont {Sharma},
  \citenamefont {Alexe},\ and\ \citenamefont {Seidel}}]{zhang-nl2021}%
  \BibitemOpen
  \bibfield  {author} {\bibinfo {author} {\bibfnamefont {Dawei}\ \bibnamefont
  {Zhang}}, \bibinfo {author} {\bibfnamefont {Zheng-Dong}\ \bibnamefont {Luo}},
  \bibinfo {author} {\bibfnamefont {Yin}\ \bibnamefont {Yao}}, \bibinfo
  {author} {\bibfnamefont {Peggy}\ \bibnamefont {Schoenherr}}, \bibinfo
  {author} {\bibfnamefont {Chuhan}\ \bibnamefont {Sha}}, \bibinfo {author}
  {\bibfnamefont {Ying}\ \bibnamefont {Pan}}, \bibinfo {author} {\bibfnamefont
  {Pankaj}\ \bibnamefont {Sharma}}, \bibinfo {author} {\bibfnamefont {Marin}\
  \bibnamefont {Alexe}}, \ and\ \bibinfo {author} {\bibfnamefont {Jan}\
  \bibnamefont {Seidel}},\ }\bibfield  {title} {\enquote {\bibinfo {title}
  {Anisotropic ion migration and electronic conduction in van der {Waals}
  ferroelectric {CuInP$_2$S$_6$}},}\ }\href {\doibase
  10.1021/acs.nanolett.0c04023} {\bibfield  {journal} {\bibinfo  {journal}
  {Nano Letters}\ }\textbf {\bibinfo {volume} {21}},\ \bibinfo {pages}
  {995--1002} (\bibinfo {year} {2021})}\BibitemShut {NoStop}%
\bibitem [{\citenamefont {Neumayer}\ \emph {et~al.}(2021)\citenamefont
  {Neumayer}, \citenamefont {Tao}, \citenamefont {O'Hara}, \citenamefont
  {Susner}, \citenamefont {McGuire}, \citenamefont {Maksymovych}, \citenamefont
  {Pantelides},\ and\ \citenamefont {Balke}}]{neumayer-aem2021}%
  \BibitemOpen
  \bibfield  {author} {\bibinfo {author} {\bibfnamefont {Sabine~M.}\
  \bibnamefont {Neumayer}}, \bibinfo {author} {\bibfnamefont {Lei}\
  \bibnamefont {Tao}}, \bibinfo {author} {\bibfnamefont {Andrew}\ \bibnamefont
  {O'Hara}}, \bibinfo {author} {\bibfnamefont {Michael~A.}\ \bibnamefont
  {Susner}}, \bibinfo {author} {\bibfnamefont {Michael~A.}\ \bibnamefont
  {McGuire}}, \bibinfo {author} {\bibfnamefont {Petro}\ \bibnamefont
  {Maksymovych}}, \bibinfo {author} {\bibfnamefont {Sokrates~T.}\ \bibnamefont
  {Pantelides}}, \ and\ \bibinfo {author} {\bibfnamefont {Nina}\ \bibnamefont
  {Balke}},\ }\bibfield  {title} {\enquote {\bibinfo {title} {The concept of
  negative capacitance in ionically conductive van der {Waals}
  ferroelectrics},}\ }\href {\doibase https://doi.org/10.1002/aenm.202103493}
  {\bibfield  {journal} {\bibinfo  {journal} {Advanced Energy Materials}\
  }\textbf {\bibinfo {volume} {11}},\ \bibinfo {pages} {2103493} (\bibinfo
  {year} {2021})}\BibitemShut {NoStop}%
\bibitem [{\citenamefont {Thouless}(1983)}]{thouless-prb1983}%
  \BibitemOpen
  \bibfield  {author} {\bibinfo {author} {\bibfnamefont {D.~J.}\ \bibnamefont
  {Thouless}},\ }\bibfield  {title} {\enquote {\bibinfo {title} {Quantization
  of particle transport},}\ }\href {\doibase 10.1103/PhysRevB.27.6083}
  {\bibfield  {journal} {\bibinfo  {journal} {Phys. Rev. B}\ }\textbf {\bibinfo
  {volume} {27}},\ \bibinfo {pages} {6083--6087} (\bibinfo {year}
  {1983})}\BibitemShut {NoStop}%
\bibitem [{\citenamefont {Jiang}\ \emph {et~al.}(2012)\citenamefont {Jiang},
  \citenamefont {Levchenko},\ and\ \citenamefont {Rappe}}]{jiang-prl2012}%
  \BibitemOpen
  \bibfield  {author} {\bibinfo {author} {\bibfnamefont {Lai}\ \bibnamefont
  {Jiang}}, \bibinfo {author} {\bibfnamefont {Sergey~V.}\ \bibnamefont
  {Levchenko}}, \ and\ \bibinfo {author} {\bibfnamefont {Andrew~M.}\
  \bibnamefont {Rappe}},\ }\bibfield  {title} {\enquote {\bibinfo {title}
  {Rigorous definition of oxidation states of ions in solids},}\ }\href
  {\doibase 10.1103/PhysRevLett.108.166403} {\bibfield  {journal} {\bibinfo
  {journal} {Phys. Rev. Lett.}\ }\textbf {\bibinfo {volume} {108}},\ \bibinfo
  {pages} {166403} (\bibinfo {year} {2012})}\BibitemShut {NoStop}%
\bibitem [{\citenamefont {Simon}\ \emph {et~al.}(1994)\citenamefont {Simon},
  \citenamefont {Ravez}, \citenamefont {Maisonneuve}, \citenamefont {Payen},\
  and\ \citenamefont {Cajipe}}]{simon-cm1994}%
  \BibitemOpen
  \bibfield  {author} {\bibinfo {author} {\bibfnamefont {A.}~\bibnamefont
  {Simon}}, \bibinfo {author} {\bibfnamefont {J.}~\bibnamefont {Ravez}},
  \bibinfo {author} {\bibfnamefont {V.}~\bibnamefont {Maisonneuve}}, \bibinfo
  {author} {\bibfnamefont {C.}~\bibnamefont {Payen}}, \ and\ \bibinfo {author}
  {\bibfnamefont {V.~B.}\ \bibnamefont {Cajipe}},\ }\bibfield  {title}
  {\enquote {\bibinfo {title} {Paraelectric-ferroelectric transition in the
  lamellar thiophosphate {CuInP$_2$S$_6$}},}\ }\href {\doibase
  10.1021/cm00045a016} {\bibfield  {journal} {\bibinfo  {journal} {Chemistry of
  Materials}\ }\textbf {\bibinfo {volume} {6}},\ \bibinfo {pages} {1575--1580}
  (\bibinfo {year} {1994})}\BibitemShut {NoStop}%
\bibitem [{\citenamefont {Maisonneuve}\ \emph {et~al.}(1995)\citenamefont
  {Maisonneuve}, \citenamefont {Evain}, \citenamefont {Payen}, \citenamefont
  {Cajipe},\ and\ \citenamefont {Molinié}}]{maisonneuve-jac1995}%
  \BibitemOpen
  \bibfield  {author} {\bibinfo {author} {\bibfnamefont {V.}~\bibnamefont
  {Maisonneuve}}, \bibinfo {author} {\bibfnamefont {M.}~\bibnamefont {Evain}},
  \bibinfo {author} {\bibfnamefont {C.}~\bibnamefont {Payen}}, \bibinfo
  {author} {\bibfnamefont {V.B.}\ \bibnamefont {Cajipe}}, \ and\ \bibinfo
  {author} {\bibfnamefont {P.}~\bibnamefont {Molinié}},\ }\bibfield  {title}
  {\enquote {\bibinfo {title} {Room-temperature crystal structure of the
  layered phase {Cu$^\text{I}$In$^\text{III}$P$_2$S$_6$}},}\ }\href {\doibase
  https://doi.org/10.1016/0925-8388(94)01416-7} {\bibfield  {journal} {\bibinfo
   {journal} {Journal of Alloys and Compounds}\ }\textbf {\bibinfo {volume}
  {218}},\ \bibinfo {pages} {157--164} (\bibinfo {year} {1995})}\BibitemShut
  {NoStop}%
\bibitem [{\citenamefont {Resta}\ and\ \citenamefont
  {Vanderbilt}(2007)}]{Resta2007}%
  \BibitemOpen
  \bibfield  {author} {\bibinfo {author} {\bibfnamefont {Raffaele}\
  \bibnamefont {Resta}}\ and\ \bibinfo {author} {\bibfnamefont {David}\
  \bibnamefont {Vanderbilt}},\ }\enquote {\bibinfo {title} {Theory of
  polarization: A modern approach},}\ in\ \href {\doibase
  10.1007/978-3-540-34591-6_2} {\emph {\bibinfo {booktitle} {Physics of
  Ferroelectrics: A Modern Perspective}}}\ (\bibinfo  {publisher} {Springer
  Berlin Heidelberg},\ \bibinfo {address} {Berlin, Heidelberg},\ \bibinfo
  {year} {2007})\ pp.\ \bibinfo {pages} {31--68}\BibitemShut {NoStop}%
\bibitem [{SM()}]{SM}%
  \BibitemOpen
  \href@noop {} {\bibinfo  {journal} {See Supplemental Material \textit{[URL to
  be inserted by publisher]} for computational details, a description of how we
  identified $\overline{M}$, and a plot of energy against $\Delta_c$ for the
  cooperative switching paths. We include files for the structures
  corresponding to the points in Fig. 3 of the main text and Fig. 1 of the
  Supplemental Material. We also include a structural animation of the cycle
  depicted in Fig. 3 of the main text}\ }\BibitemShut {NoStop}%
\bibitem [{\citenamefont {Henkelman}\ and\ \citenamefont
  {Jónsson}(1999)}]{dimer1-jcp1999}%
  \BibitemOpen
\bibfield  {journal} {  }\bibfield  {author} {\bibinfo {author} {\bibfnamefont
  {Graeme}\ \bibnamefont {Henkelman}}\ and\ \bibinfo {author} {\bibfnamefont
  {Hannes}\ \bibnamefont {Jónsson}},\ }\bibfield  {title} {\enquote {\bibinfo
  {title} {A dimer method for finding saddle points on high dimensional
  potential surfaces using only first derivatives},}\ }\href {\doibase
  10.1063/1.480097} {\bibfield  {journal} {\bibinfo  {journal} {The Journal of
  Chemical Physics}\ }\textbf {\bibinfo {volume} {111}},\ \bibinfo {pages}
  {7010--7022} (\bibinfo {year} {1999})}\BibitemShut {NoStop}%
\bibitem [{\citenamefont {Heyden}\ \emph {et~al.}(2005)\citenamefont {Heyden},
  \citenamefont {Bell},\ and\ \citenamefont {Keil}}]{dimer2-jcp2005}%
  \BibitemOpen
  \bibfield  {author} {\bibinfo {author} {\bibfnamefont {Andreas}\ \bibnamefont
  {Heyden}}, \bibinfo {author} {\bibfnamefont {Alexis~T.}\ \bibnamefont
  {Bell}}, \ and\ \bibinfo {author} {\bibfnamefont {Frerich~J.}\ \bibnamefont
  {Keil}},\ }\bibfield  {title} {\enquote {\bibinfo {title} {Efficient methods
  for finding transition states in chemical reactions: Comparison of improved
  dimer method and partitioned rational function optimization method},}\ }\href
  {\doibase 10.1063/1.2104507} {\bibfield  {journal} {\bibinfo  {journal} {The
  Journal of Chemical Physics}\ }\textbf {\bibinfo {volume} {123}},\ \bibinfo
  {pages} {224101} (\bibinfo {year} {2005})}\BibitemShut {NoStop}%
\bibitem [{\citenamefont {Di\'eguez}\ and\ \citenamefont
  {Vanderbilt}(2006)}]{dieguez-prl2006}%
  \BibitemOpen
  \bibfield  {author} {\bibinfo {author} {\bibfnamefont {Oswaldo}\ \bibnamefont
  {Di\'eguez}}\ and\ \bibinfo {author} {\bibfnamefont {David}\ \bibnamefont
  {Vanderbilt}},\ }\bibfield  {title} {\enquote {\bibinfo {title}
  {First-principles calculations for insulators at constant polarization},}\
  }\href {\doibase 10.1103/PhysRevLett.96.056401} {\bibfield  {journal}
  {\bibinfo  {journal} {Phys. Rev. Lett.}\ }\textbf {\bibinfo {volume} {96}},\
  \bibinfo {pages} {056401} (\bibinfo {year} {2006})}\BibitemShut {NoStop}%
\bibitem [{\citenamefont {Stengel}\ \emph
  {et~al.}(2009{\natexlab{a}})\citenamefont {Stengel}, \citenamefont
  {Spaldin},\ and\ \citenamefont {Vanderbilt}}]{stengel-natphys2009}%
  \BibitemOpen
  \bibfield  {author} {\bibinfo {author} {\bibfnamefont {Massimiliano}\
  \bibnamefont {Stengel}}, \bibinfo {author} {\bibfnamefont {Nicola~A.}\
  \bibnamefont {Spaldin}}, \ and\ \bibinfo {author} {\bibfnamefont {David}\
  \bibnamefont {Vanderbilt}},\ }\bibfield  {title} {\enquote {\bibinfo {title}
  {Electric displacement as the fundamental variable in electronic-structure
  calculations},}\ }\href {\doibase 10.1038/nphys1185} {\bibfield  {journal}
  {\bibinfo  {journal} {Nature Physics}\ }\textbf {\bibinfo {volume} {5}},\
  \bibinfo {pages} {304--308} (\bibinfo {year}
  {2009}{\natexlab{a}})}\BibitemShut {NoStop}%
\bibitem [{\citenamefont {Stengel}\ \emph
  {et~al.}(2009{\natexlab{b}})\citenamefont {Stengel}, \citenamefont
  {Vanderbilt},\ and\ \citenamefont {Spaldin}}]{stengel-prb2009}%
  \BibitemOpen
  \bibfield  {author} {\bibinfo {author} {\bibfnamefont {Massimiliano}\
  \bibnamefont {Stengel}}, \bibinfo {author} {\bibfnamefont {David}\
  \bibnamefont {Vanderbilt}}, \ and\ \bibinfo {author} {\bibfnamefont
  {Nicola~A.}\ \bibnamefont {Spaldin}},\ }\bibfield  {title} {\enquote
  {\bibinfo {title} {First-principles modeling of ferroelectric capacitors via
  constrained displacement field calculations},}\ }\href {\doibase
  10.1103/PhysRevB.80.224110} {\bibfield  {journal} {\bibinfo  {journal} {Phys.
  Rev. B}\ }\textbf {\bibinfo {volume} {80}},\ \bibinfo {pages} {224110}
  (\bibinfo {year} {2009}{\natexlab{b}})}\BibitemShut {NoStop}%
\bibitem [{\citenamefont {Habbal}\ \emph {et~al.}(1978)\citenamefont {Habbal},
  \citenamefont {Zvirgzds},\ and\ \citenamefont {Scott}}]{habbal-jcp1978}%
  \BibitemOpen
  \bibfield  {author} {\bibinfo {author} {\bibfnamefont {F.}~\bibnamefont
  {Habbal}}, \bibinfo {author} {\bibfnamefont {J.~A.}\ \bibnamefont
  {Zvirgzds}}, \ and\ \bibinfo {author} {\bibfnamefont {J.~F.}\ \bibnamefont
  {Scott}},\ }\bibfield  {title} {\enquote {\bibinfo {title} {Raman
  spectroscopy of structural phase transitions in
  {Ag$_{26}$I$_{18}$W$_4$O$_{16}$}},}\ }\href {\doibase 10.1063/1.436487}
  {\bibfield  {journal} {\bibinfo  {journal} {The Journal of Chemical Physics}\
  }\textbf {\bibinfo {volume} {69}},\ \bibinfo {pages} {4984--4989} (\bibinfo
  {year} {1978})}\BibitemShut {NoStop}%
\bibitem [{\citenamefont {Scott}\ \emph {et~al.}(1980)\citenamefont {Scott},
  \citenamefont {Habbal},\ and\ \citenamefont {Zvirgzds}}]{scott-jcp1980}%
  \BibitemOpen
  \bibfield  {author} {\bibinfo {author} {\bibfnamefont {J.~F.}\ \bibnamefont
  {Scott}}, \bibinfo {author} {\bibfnamefont {F.}~\bibnamefont {Habbal}}, \
  and\ \bibinfo {author} {\bibfnamefont {J.~A.}\ \bibnamefont {Zvirgzds}},\
  }\bibfield  {title} {\enquote {\bibinfo {title} {Ferroelectric phase
  transition in the superionic conductor {Ag$_{26}$I$_{18}$W$_4$O$_{16}$}},}\
  }\href {\doibase 10.1063/1.439424} {\bibfield  {journal} {\bibinfo  {journal}
  {The Journal of Chemical Physics}\ }\textbf {\bibinfo {volume} {72}},\
  \bibinfo {pages} {2760--2762} (\bibinfo {year} {1980})}\BibitemShut {NoStop}%
\bibitem [{\citenamefont {Hoshino}\ \emph {et~al.}(1981)\citenamefont
  {Hoshino}, \citenamefont {Fujishita}, \citenamefont {Takashige},\ and\
  \citenamefont {Sakuma}}]{hoshino-ssi1981}%
  \BibitemOpen
  \bibfield  {author} {\bibinfo {author} {\bibfnamefont {S.}~\bibnamefont
  {Hoshino}}, \bibinfo {author} {\bibfnamefont {H.}~\bibnamefont {Fujishita}},
  \bibinfo {author} {\bibfnamefont {M.}~\bibnamefont {Takashige}}, \ and\
  \bibinfo {author} {\bibfnamefont {T.}~\bibnamefont {Sakuma}},\ }\bibfield
  {title} {\enquote {\bibinfo {title} {Phase transition of {Ag$_3$SX} ({X=I,
  Br})},}\ }\href {\doibase https://doi.org/10.1016/0167-2738(81)90050-3}
  {\bibfield  {journal} {\bibinfo  {journal} {Solid State Ionics}\ }\textbf
  {\bibinfo {volume} {3-4}},\ \bibinfo {pages} {35--39} (\bibinfo {year}
  {1981})}\BibitemShut {NoStop}%
\bibitem [{\citenamefont {Stefanovich}\ \emph {et~al.}(1984)\citenamefont
  {Stefanovich}, \citenamefont {Kalinin},\ and\ \citenamefont
  {Nogai}}]{stefanovich-FE1984}%
  \BibitemOpen
  \bibfield  {author} {\bibinfo {author} {\bibfnamefont {S.~Yu.}\ \bibnamefont
  {Stefanovich}}, \bibinfo {author} {\bibfnamefont {V.~B.}\ \bibnamefont
  {Kalinin}}, \ and\ \bibinfo {author} {\bibfnamefont {A.}~\bibnamefont
  {Nogai}},\ }\bibfield  {title} {\enquote {\bibinfo {title}
  {Ferroelectric-superionic conductor phase transitions in
  {Na$_3$Sc$_2$(PO$_4$)$_3$} and {ITS} isomorphes},}\ }\href {\doibase
  10.1080/00150198408015400} {\bibfield  {journal} {\bibinfo  {journal}
  {Ferroelectrics}\ }\textbf {\bibinfo {volume} {55}},\ \bibinfo {pages}
  {325--328} (\bibinfo {year} {1984})}\BibitemShut {NoStop}%
\bibitem [{\citenamefont {Stefanovich}\ \emph {et~al.}(1985)\citenamefont
  {Stefanovich}, \citenamefont {Yanovsky}, \citenamefont {Astafyev},
  \citenamefont {Voronkova},\ and\ \citenamefont
  {Venevtsev}}]{stefanovich-jjap1985}%
  \BibitemOpen
  \bibfield  {author} {\bibinfo {author} {\bibfnamefont {S.~Yu.}\ \bibnamefont
  {Stefanovich}}, \bibinfo {author} {\bibfnamefont {V.~K.}\ \bibnamefont
  {Yanovsky}}, \bibinfo {author} {\bibfnamefont {A.~V.}\ \bibnamefont
  {Astafyev}}, \bibinfo {author} {\bibfnamefont {V.~I.}\ \bibnamefont
  {Voronkova}}, \ and\ \bibinfo {author} {\bibfnamefont {Yu.~N.}\ \bibnamefont
  {Venevtsev}},\ }\bibfield  {title} {\enquote {\bibinfo {title}
  {Ferroelectric-superionic conductor phase transitions in crystals
  {MeNbWO$_6$·nH$_2$O} ({Me=Tl, Rb})},}\ }\href {\doibase
  10.7567/JJAPS.24S2.373} {\bibfield  {journal} {\bibinfo  {journal} {Japanese
  Journal of Applied Physics}\ }\textbf {\bibinfo {volume} {24}},\ \bibinfo
  {pages} {373} (\bibinfo {year} {1985})}\BibitemShut {NoStop}%
\bibitem [{\citenamefont {Baranov}\ \emph {et~al.}(1989)\citenamefont
  {Baranov}, \citenamefont {Khiznichenko},\ and\ \citenamefont
  {Shuvalov}}]{baranov-FE1989}%
  \BibitemOpen
  \bibfield  {author} {\bibinfo {author} {\bibfnamefont {A.~I.}\ \bibnamefont
  {Baranov}}, \bibinfo {author} {\bibfnamefont {V.~P.}\ \bibnamefont
  {Khiznichenko}}, \ and\ \bibinfo {author} {\bibfnamefont {L.~A.}\
  \bibnamefont {Shuvalov}},\ }\bibfield  {title} {\enquote {\bibinfo {title}
  {High temperature phase transitions and proton conductivity in some
  kdp-family crystals},}\ }\href {\doibase 10.1080/00150198908007907}
  {\bibfield  {journal} {\bibinfo  {journal} {Ferroelectrics}\ }\textbf
  {\bibinfo {volume} {100}},\ \bibinfo {pages} {135--141} (\bibinfo {year}
  {1989})}\BibitemShut {NoStop}%
\bibitem [{\citenamefont {Scott}(1999)}]{scott-ssi1999}%
  \BibitemOpen
  \bibfield  {author} {\bibinfo {author} {\bibfnamefont {J.F.}\ \bibnamefont
  {Scott}},\ }\bibfield  {title} {\enquote {\bibinfo {title} {A comparison of
  {Ag}- and proton-conducting ferroelectrics},}\ }\href {\doibase
  https://doi.org/10.1016/S0167-2738(99)00168-X} {\bibfield  {journal}
  {\bibinfo  {journal} {Solid State Ionics}\ }\textbf {\bibinfo {volume}
  {125}},\ \bibinfo {pages} {141--146} (\bibinfo {year} {1999})}\BibitemShut
  {NoStop}%
\bibitem [{\citenamefont {Neumayer}\ \emph {et~al.}(2022)\citenamefont
  {Neumayer}, \citenamefont {Si}, \citenamefont {Li}, \citenamefont {Liao},
  \citenamefont {Tao}, \citenamefont {O’Hara}, \citenamefont {Pantelides},
  \citenamefont {Ye}, \citenamefont {Maksymovych},\ and\ \citenamefont
  {Balke}}]{neumayer-ami2022}%
  \BibitemOpen
  \bibfield  {author} {\bibinfo {author} {\bibfnamefont {Sabine~M.}\
  \bibnamefont {Neumayer}}, \bibinfo {author} {\bibfnamefont {Mengwei}\
  \bibnamefont {Si}}, \bibinfo {author} {\bibfnamefont {Junkang}\ \bibnamefont
  {Li}}, \bibinfo {author} {\bibfnamefont {Pai-Ying}\ \bibnamefont {Liao}},
  \bibinfo {author} {\bibfnamefont {Lei}\ \bibnamefont {Tao}}, \bibinfo
  {author} {\bibfnamefont {Andrew}\ \bibnamefont {O’Hara}}, \bibinfo {author}
  {\bibfnamefont {Sokrates~T.}\ \bibnamefont {Pantelides}}, \bibinfo {author}
  {\bibfnamefont {Peide~D.}\ \bibnamefont {Ye}}, \bibinfo {author}
  {\bibfnamefont {Petro}\ \bibnamefont {Maksymovych}}, \ and\ \bibinfo {author}
  {\bibfnamefont {Nina}\ \bibnamefont {Balke}},\ }\bibfield  {title} {\enquote
  {\bibinfo {title} {Ionic control over ferroelectricity in {2D} layered van
  der {Waals} capacitors},}\ }\href {\doibase 10.1021/acsami.1c18683}
  {\bibfield  {journal} {\bibinfo  {journal} {ACS Applied Materials \&
  Interfaces}\ }\textbf {\bibinfo {volume} {14}},\ \bibinfo {pages}
  {3018--3026} (\bibinfo {year} {2022})}\BibitemShut {NoStop}%
\bibitem [{\citenamefont {Lindgren}\ \emph {et~al.}(2018)\citenamefont
  {Lindgren}, \citenamefont {Ievlev}, \citenamefont {Jesse}, \citenamefont
  {Ovchinnikova}, \citenamefont {Kalinin}, \citenamefont {Vasudevan},\ and\
  \citenamefont {Canalias}}]{lindgren-ami2018}%
  \BibitemOpen
  \bibfield  {author} {\bibinfo {author} {\bibfnamefont {Gustav}\ \bibnamefont
  {Lindgren}}, \bibinfo {author} {\bibfnamefont {Anton}\ \bibnamefont
  {Ievlev}}, \bibinfo {author} {\bibfnamefont {Stephen}\ \bibnamefont {Jesse}},
  \bibinfo {author} {\bibfnamefont {Olga~S.}\ \bibnamefont {Ovchinnikova}},
  \bibinfo {author} {\bibfnamefont {Sergei~V.}\ \bibnamefont {Kalinin}},
  \bibinfo {author} {\bibfnamefont {Rama~K.}\ \bibnamefont {Vasudevan}}, \ and\
  \bibinfo {author} {\bibfnamefont {Carlota}\ \bibnamefont {Canalias}},\
  }\bibfield  {title} {\enquote {\bibinfo {title} {Elasticity modulation due to
  polarization reversal and ionic motion in the ferroelectric superionic
  conductor {KTiOPO$_4$}},}\ }\href {\doibase 10.1021/acsami.8b07537}
  {\bibfield  {journal} {\bibinfo  {journal} {ACS Applied Materials \&
  Interfaces}\ }\textbf {\bibinfo {volume} {10}},\ \bibinfo {pages}
  {32298--32303} (\bibinfo {year} {2018})}\BibitemShut {NoStop}%
\bibitem [{\citenamefont {Maglione}\ \emph {et~al.}(2016)\citenamefont
  {Maglione}, \citenamefont {Theerthan}, \citenamefont {Rodriguez},
  \citenamefont {Peña}, \citenamefont {Canalias}, \citenamefont {Ménaert},\
  and\ \citenamefont {Boulanger}}]{maglione-ome2016}%
  \BibitemOpen
  \bibfield  {author} {\bibinfo {author} {\bibfnamefont {Mario}\ \bibnamefont
  {Maglione}}, \bibinfo {author} {\bibfnamefont {Anand}\ \bibnamefont
  {Theerthan}}, \bibinfo {author} {\bibfnamefont {Vincent}\ \bibnamefont
  {Rodriguez}}, \bibinfo {author} {\bibfnamefont {Alexandra}\ \bibnamefont
  {Peña}}, \bibinfo {author} {\bibfnamefont {Carlota}\ \bibnamefont
  {Canalias}}, \bibinfo {author} {\bibfnamefont {Bertrand}\ \bibnamefont
  {Ménaert}}, \ and\ \bibinfo {author} {\bibfnamefont {Benoît}\ \bibnamefont
  {Boulanger}},\ }\bibfield  {title} {\enquote {\bibinfo {title} {Intrinsic
  ionic screening of the ferroelectric polarization of {KTP} revealed by
  second-harmonic generation microscopy},}\ }\href {\doibase
  10.1364/OME.6.000137} {\bibfield  {journal} {\bibinfo  {journal} {Opt. Mater.
  Express}\ }\textbf {\bibinfo {volume} {6}},\ \bibinfo {pages} {137--145}
  (\bibinfo {year} {2016})}\BibitemShut {NoStop}%
\end{thebibliography}%


\begin{thebibliography}{8}%
\makeatletter
\providecommand \@ifxundefined [1]{%
 \@ifx{#1\undefined}
}%
\providecommand \@ifnum [1]{%
 \ifnum #1\expandafter \@firstoftwo
 \else \expandafter \@secondoftwo
 \fi
}%
\providecommand \@ifx [1]{%
 \ifx #1\expandafter \@firstoftwo
 \else \expandafter \@secondoftwo
 \fi
}%
\providecommand \natexlab [1]{#1}%
\providecommand \enquote  [1]{``#1''}%
\providecommand \bibnamefont  [1]{#1}%
\providecommand \bibfnamefont [1]{#1}%
\providecommand \citenamefont [1]{#1}%
\providecommand \href@noop [0]{\@secondoftwo}%
\providecommand \href [0]{\begingroup \@sanitize@url \@href}%
\providecommand \@href[1]{\@@startlink{#1}\@@href}%
\providecommand \@@href[1]{\endgroup#1\@@endlink}%
\providecommand \@sanitize@url [0]{\catcode `\\12\catcode `\$12\catcode
  `\&12\catcode `\#12\catcode `\^12\catcode `\_12\catcode `\%12\relax}%
\providecommand \@@startlink[1]{}%
\providecommand \@@endlink[0]{}%
\providecommand \url  [0]{\begingroup\@sanitize@url \@url }%
\providecommand \@url [1]{\endgroup\@href {#1}{\urlprefix }}%
\providecommand \urlprefix  [0]{URL }%
\providecommand \Eprint [0]{\href }%
\providecommand \doibase [0]{http://dx.doi.org/}%
\providecommand \selectlanguage [0]{\@gobble}%
\providecommand \bibinfo  [0]{\@secondoftwo}%
\providecommand \bibfield  [0]{\@secondoftwo}%
\providecommand \translation [1]{[#1]}%
\providecommand \BibitemOpen [0]{}%
\providecommand \bibitemStop [0]{}%
\providecommand \bibitemNoStop [0]{.\EOS\space}%
\providecommand \EOS [0]{\spacefactor3000\relax}%
\providecommand \BibitemShut  [1]{\csname bibitem#1\endcsname}%
\let\auto@bib@innerbib\@empty
\bibitem [{\citenamefont {Kresse}\ and\ \citenamefont
  {Furthmüller}(1996)}]{kresse-prb1996}%
  \BibitemOpen
  \bibfield  {author} {\bibinfo {author} {\bibfnamefont {G.}~\bibnamefont
  {Kresse}}\ and\ \bibinfo {author} {\bibfnamefont {J.}~\bibnamefont
  {Furthmüller}},\ }\href {\doibase 10.1103/PhysRevB.54.11169} {\bibfield
  {journal} {\bibinfo  {journal} {Phys. Rev. B}\ }\textbf {\bibinfo {volume}
  {54}},\ \bibinfo {pages} {11169} (\bibinfo {year} {1996})}\BibitemShut
  {NoStop}%
\bibitem [{\citenamefont {Kresse}\ and\ \citenamefont
  {Joubert}(1999)}]{kresse-prb1999}%
  \BibitemOpen
  \bibfield  {author} {\bibinfo {author} {\bibfnamefont {G.}~\bibnamefont
  {Kresse}}\ and\ \bibinfo {author} {\bibfnamefont {D.}~\bibnamefont
  {Joubert}},\ }\href {\doibase 10.1103/PhysRevB.59.1758} {\bibfield  {journal}
  {\bibinfo  {journal} {Phys. Rev. B}\ }\textbf {\bibinfo {volume} {59}},\
  \bibinfo {pages} {1758} (\bibinfo {year} {1999})}\BibitemShut {NoStop}%
\bibitem [{\citenamefont {Perdew}\ \emph {et~al.}(1996)\citenamefont {Perdew},
  \citenamefont {Burke},\ and\ \citenamefont {Ernzerhof}}]{pbe-prl1997}%
  \BibitemOpen
  \bibfield  {author} {\bibinfo {author} {\bibfnamefont {J.~P.}\ \bibnamefont
  {Perdew}}, \bibinfo {author} {\bibfnamefont {K.}~\bibnamefont {Burke}}, \
  and\ \bibinfo {author} {\bibfnamefont {M.}~\bibnamefont {Ernzerhof}},\ }\href
  {\doibase 10.1103/PhysRevLett.77.3865} {\bibfield  {journal} {\bibinfo
  {journal} {Phys. Rev. Lett.}\ }\textbf {\bibinfo {volume} {77}},\ \bibinfo
  {pages} {3865} (\bibinfo {year} {1996})}\BibitemShut {NoStop}%
\bibitem [{\citenamefont {Grimme}\ \emph {et~al.}(2010)\citenamefont {Grimme},
  \citenamefont {Antony}, \citenamefont {Ehrlich},\ and\ \citenamefont
  {Krieg}}]{d3-jcp2010}%
  \BibitemOpen
  \bibfield  {author} {\bibinfo {author} {\bibfnamefont {S.}~\bibnamefont
  {Grimme}}, \bibinfo {author} {\bibfnamefont {J.}~\bibnamefont {Antony}},
  \bibinfo {author} {\bibfnamefont {S.}~\bibnamefont {Ehrlich}}, \ and\
  \bibinfo {author} {\bibfnamefont {H.}~\bibnamefont {Krieg}},\ }\href
  {\doibase 10.1063/1.3382344} {\bibfield  {journal} {\bibinfo  {journal} {The
  Journal of Chemical Physics}\ }\textbf {\bibinfo {volume} {132}},\ \bibinfo
  {pages} {154104} (\bibinfo {year} {2010})}\BibitemShut {NoStop}%
\bibitem [{\citenamefont {Mills}\ and\ \citenamefont
  {J\'onsson}(1994)}]{mills-prl1994}%
  \BibitemOpen
  \bibfield  {author} {\bibinfo {author} {\bibfnamefont {G.}~\bibnamefont
  {Mills}}\ and\ \bibinfo {author} {\bibfnamefont {H.}~\bibnamefont
  {J\'onsson}},\ }\href {\doibase 10.1103/PhysRevLett.72.1124} {\bibfield
  {journal} {\bibinfo  {journal} {Phys. Rev. Lett.}\ }\textbf {\bibinfo
  {volume} {72}},\ \bibinfo {pages} {1124} (\bibinfo {year}
  {1994})}\BibitemShut {NoStop}%
\bibitem [{\citenamefont {Mills}\ \emph {et~al.}(1995)\citenamefont {Mills},
  \citenamefont {Jónsson},\ and\ \citenamefont
  {Schenter}}]{mills-surfsci1995}%
  \BibitemOpen
  \bibfield  {author} {\bibinfo {author} {\bibfnamefont {G.}~\bibnamefont
  {Mills}}, \bibinfo {author} {\bibfnamefont {H.}~\bibnamefont {Jónsson}}, \
  and\ \bibinfo {author} {\bibfnamefont {G.~K.}\ \bibnamefont {Schenter}},\
  }\href {\doibase https://doi.org/10.1016/0039-6028(94)00731-4} {\bibfield
  {journal} {\bibinfo  {journal} {Surface Science}\ }\textbf {\bibinfo {volume}
  {324}},\ \bibinfo {pages} {305} (\bibinfo {year} {1995})}\BibitemShut
  {NoStop}%
\bibitem [{\citenamefont {Jónsson}\ \emph {et~al.}()\citenamefont {Jónsson},
  \citenamefont {Mills},\ and\ \citenamefont {Jacobsen}}]{jonsson-pneb1998}%
  \BibitemOpen
  \bibfield  {author} {\bibinfo {author} {\bibfnamefont {H.}~\bibnamefont
  {Jónsson}}, \bibinfo {author} {\bibfnamefont {G.}~\bibnamefont {Mills}}, \
  and\ \bibinfo {author} {\bibfnamefont {K.~W.}\ \bibnamefont {Jacobsen}},\
  }\enquote {\bibinfo {title} {Nudged elastic band method for finding minimum
  energy paths of transitions},}\ in\ \href {\doibase
  10.1142/9789812839664_0016} {\emph {\bibinfo {booktitle} {Classical and
  Quantum Dynamics in Condensed Phase Simulations}}},\ pp.\ \bibinfo {pages}
  {385--404}\BibitemShut {NoStop}%
\bibitem [{\citenamefont {King-Smith}\ and\ \citenamefont
  {Vanderbilt}(1993)}]{kingsmith-prb1993}%
  \BibitemOpen
  \bibfield  {author} {\bibinfo {author} {\bibfnamefont {R.~D.}\ \bibnamefont
  {King-Smith}}\ and\ \bibinfo {author} {\bibfnamefont {D.}~\bibnamefont
  {Vanderbilt}},\ }\href {\doibase 10.1103/PhysRevB.47.1651} {\bibfield
  {journal} {\bibinfo  {journal} {Phys. Rev. B}\ }\textbf {\bibinfo {volume}
  {47}},\ \bibinfo {pages} {1651} (\bibinfo {year} {1993})}\BibitemShut
  {NoStop}%
\end{thebibliography}%

\end{document}



\title{Supplemental Material: Cyclic Ferroelectric Switching and Quantized Charge Transport in CuInP$_2$S$_6$}

\author{Daniel Seleznev}
\affiliation{Department of Physics and Astronomy, Center for Materials Theory, Rutgers University, Piscataway, New Jersey 08854-8019, USA}

\author{Sobhit Singh}
\affiliation{Department of Mechanical Engineering, University of Rochester, Rochester, New York 14627, USA}

\author{John Bonini}
\affiliation{Center for Computational Quantum Physics, Flatiron Institute, 162 5th Avenue, New York, New York 10010, USA}

\author{Karin M. Rabe}
\affiliation{Department of Physics and Astronomy, Center for Materials Theory, Rutgers University, Piscataway, New Jersey 08854-8019, USA}

\author{David Vanderbilt}
\affiliation{Department of Physics and Astronomy, Center for Materials Theory, Rutgers University, Piscataway, New Jersey 08854-8019, USA}


{\let\clearpage\relax\maketitle\vspace{-15ex}}

In this Supplemental Material, we provide computational details on all of our calculations, as well as discuss how we identified the midpoint structure $\overline{M}$ of the cross-gap cooperative switching path. We additionally provide an energy profile for the aforementioned switching path.

\section{Computational details}
All first-principles density functional theory (DFT) calculations are performed using the Vienna Ab-initio Simulation Package (VASP) with the projector augmented wave (PAW) method \cite{kresse-prb1996,kresse-prb1999}, and employ the generalized-gradient approximation to the exchange-correlation functional, as parametrized by Perdew-Burke-Ernzerhof \cite{pbe-prl1997}. van der Waals (vdW) interactions are taken into account by using the zero-damping D3 method of Grimme \cite{d3-jcp2010}. The energy convergence criterion for all self-consistent DFT calculations is set to $10^{-7}$\,eV, while the force convergence criterion for atomic coordinate relaxation is set at $10^{-3}\,\text{eV/\AA}$; for plain elastic band (PEB) and nudged elastic band (NEB) calculations \cite{mills-prl1994,mills-surfsci1995,jonsson-pneb1998}, the latter is set to $0.025\,\text{eV/\AA}$. All calculations employ a kinetic energy cutoff of 400\,eV for the plane waves, and all employ $\Gamma$-centered $k$-meshes. For electric polarization (computed using the Berry phase theory \cite{kingsmith-prb1993}) the $k$-mesh size is $8\times8\times15$, while for all other calculations it is $8\times8\times4$. All calculations are performed for systems at zero temperature.

\section{Midpoint of the through-gap cooperative switching path}
As discussed in the main text, despite being found at the same value of $\Delta_c=0.25$, the $M^+$ and $M^-$ structures are distinct, and therefore do not link up to form a continuous switching path. To remedy this problem, we turn to the PEB and NEB methods to identify a path connecting the two structures. In all PEB and NEB calculations about to be described, the lattice vectors and internal atomic coordinates are permitted to vary so as to preserve the symmetry of the initial paths provided. The initial paths are obtained by linearly interpolating the lattice vectors and internal atomic coordinates between the two endpoint structures of the path.

Initially, we perform a PEB calculation initialized on a path with a single structure between $M^+$ and $M^-$. The structure is obtained by selecting the midpoint of the linear interpolation between $M^+$ and $M^-$, and is found in space group $C2/c$. The output of this PEB calculation is subsequently used as the only input intermediate structure in an NEB calculation with the same anchor points; we label the result of this follow-up NEB calculation as $\widetilde{M}$. $\widetilde{M}$ is then used to initialize an NEB calculation in which the initial path is composed of structures residing on the combination of linearly interpolated paths between $M^+$ and $\widetilde{M}$, and $\widetilde{M}$ and $M^-$. Although the calculation satisfies the convergence criteria, we find that resulting path is still discontinuous.  

We subsequently change the anchor points of the NEB calculation, selecting the states at $\Delta_c=0.18$ and $0.32$; under appropriate applied strain these states become the $\pm$HP states residing on the quadruple-well potential for CIPS. Using the new anchor points and selecting the structure residing on the midpoint of the linear interpolation between them, we perform the same sequence of PEB followed by NEB calculation as described above. The output is labeled $\widehat{M}$. We then generate an initial path composed of structures residing on the linearly interpolated paths between the structure at $\Delta_c=0.18$ and $\widehat{M}$, and $\widehat{M}$ and the structure at $\Delta_c=0.32$, respectively. In the course of the calculation, $\widehat{M}$ evolves into the structure we refer to as $\overline{M}$, found in space group $C2/c$. We find that $\overline{M}$ does not differ significantly from $\widetilde{M}$, and we thus identify the former as defining the energy barrier for cooperative cross-gap switching.

\newpage

\section{Energy profile of the cooperative switching path}  
\begin{figure}[h]
\centering
\includegraphics[width= 15.6cm]{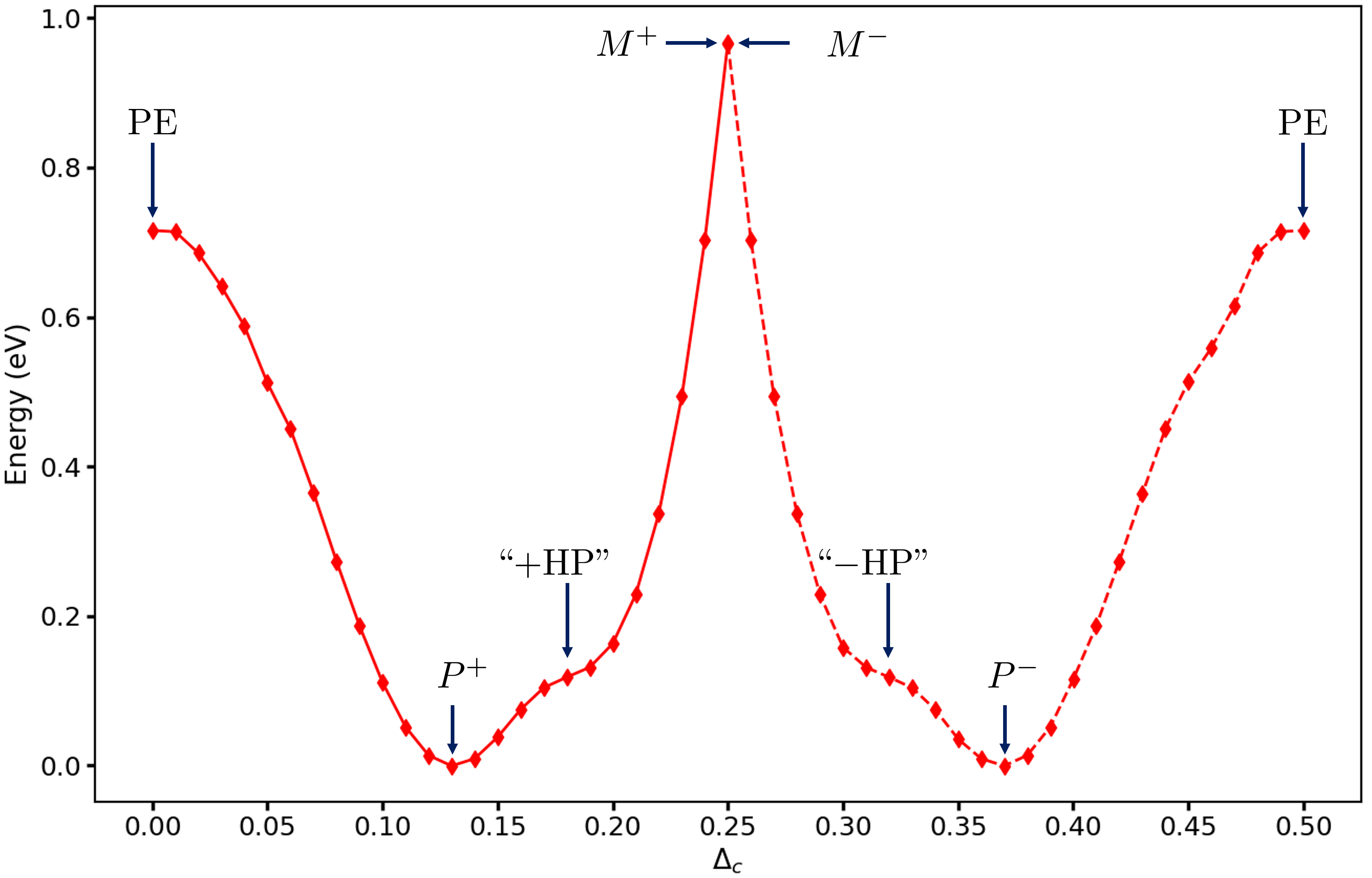}
\caption{The energy profile as a function of $\Delta_c$ of the combined through-layer and cross-gap cooperative switching paths. We indicate the locations of the points corresponding to the PE phase, $P^{\pm}$, and $M^{\pm}$ structures. By ``$\pm$HP" we indicate the structures that under applied strain will turn into the proper $\pm$HP states residing on the quadruple-well potential of CIPS. Although in the main text the states derived from $P^-$ feature negative values of $\Delta_c$, here we have used the fact that $\Delta_c$ is defined modulo $1/2$ to translate the corresponding portion of the curve to lie in the range $0.25\leq\Delta_c\leq0.5$.}
\label{fig:supfig1}
\end{figure}

\bibliography{bibfile.bib}